\documentclass[iop]{emulateapj}
\usepackage[breaklinks,colorlinks,citecolor=blue]{hyperref}
\usepackage{xspace,bold-extra}

\newcommand{\Lprime}[3]{L$^{\prime}_{\textnormal{\scriptsize{#1 (#2--#3)}}}$\xspace}
\newcommand{\CO}[2]{CO~(#1--#2)\xspace}
\newcommand{\HCN}[2]{HCN~(#1--#2)\xspace}
\newcommand{\HCO}[2]{HCO$^+$~(#1--#2)\xspace}
\newcommand{\HNC}[2]{HNC~(#1--#2)\xspace}

\newcommand{\HCNj}[2]{HCN~(J~=~$#1\rightarrow#2$)\xspace}
\newcommand{\HCOj}[2]{HCO$^+$~(J~=~$#1\rightarrow#2$)\xspace}
\newcommand{\HNCj}[2]{HNC~(J~=~$#1\rightarrow#2$)\xspace}
\newcommand{\CCHn}[2]{CCH~(N~=~$#1\rightarrow#2$)\xspace}
\newcommand{\cii}{[C~\textsc{ii}]\xspace}
\newcommand{\LIR}{L$_{\textnormal{\scriptsize{IR}}}$\xspace}
\newcommand{\LIRr}{L$_{\textnormal{\scriptsize{IR}}}[8-1000~\mu m]$\xspace}
\newcommand{\LFIR}{L$_{\textnormal{\scriptsize{FIR}}}$\xspace}
\newcommand{\LFIRr}{L$_{\textnormal{\scriptsize{FIR}}}[40-400~\mu m]$\xspace}
\newcommand{\cmc}{cm$^{-3}$\xspace}
\newcommand{\cmcol}{cm$^{-2}$\xspace}
\newcommand{\kms}{km~s$^{-1}$\xspace}
\newcommand{\Lsun}{L$_{\odot}$\xspace}

\newcommand{\myemail}{gprivon@astro-udec.cl}

\slugcomment{Accepted for publication in ApJ.}

\shorttitle{\HCN{1}{0} and \HCO{1}{0} in GOALS (U)LIRGs}
\shortauthors{Privon et al.}

\begin{document}

\title{Excitation Mechanisms for \HCN{1}{0} and \HCO{1}{0} in Galaxies from the Great Observatories All-sky LIRG Survey\footnotemark[1]}
\footnotetext[1]{Based on observations carried out with the IRAM 30m Telescope.  IRAM is supported by INSU/CNRS (France), MPG (Germany) and IGN (Spain).}
\author{G. C. Privon\altaffilmark{2,}\altaffilmark{3.}\altaffilmark{4}}
\author{R. Herrero-Illana\altaffilmark{5}}
\author{A. S. Evans\altaffilmark{2,}\altaffilmark{6}}
\author{K. Iwasawa\altaffilmark{7}}
\author{M. A. Perez-Torres\altaffilmark{5}}
\author{L. Armus\altaffilmark{8}}
\author{T. D{\' i}az-Santos\altaffilmark{8,}\altaffilmark{9}}
\author{E. J. Murphy\altaffilmark{10}}
\author{S. Stierwalt\altaffilmark{2}}
\author{S. Aalto\altaffilmark{11}}
\author{J. M. Mazzarella\altaffilmark{10}}
\author{L. Barcos-Mu{\~ n}oz\altaffilmark{2}}
\author{H. J. Borish\altaffilmark{2}}
\author{H. Inami\altaffilmark{13}}
\author{D.-C. Kim\altaffilmark{6}}
\author{E. Treister\altaffilmark{3}}
\author{J. A. Surace\altaffilmark{8}}
\author{S. Lord\altaffilmark{14}}
\author{J. Conway\altaffilmark{11}}
\author{D. T. Frayer\altaffilmark{12}}
\author{A. Alberdi\altaffilmark{5}}
\email{\myemail}

\altaffiltext{2}{Department of Astronomy, University of Virginia, Charlottesville, VA, USA}
\altaffiltext{3}{Departamento de Astronom{\' i}a, Universidad de Concepci{\' o}n, Concepci{\' o}n, Chile}
\altaffiltext{4}{Visiting Graduate Student Research Fellow (2013), NASA Infrared Processing and Analysis Center, California Institute of Technology, Pasadena, CA, USA}
\altaffiltext{5}{Instituto de Astrof{\' i}sica de Andaluc{\' i}aa-CSIC, Glorieta de la Astronom\'ia s/n, 18008, Granada, Spain}
\altaffiltext{6}{National Radio Astronomy Observatory, Charlottesville, VA, USA}
\altaffiltext{7}{ ICREA and Institut de Ci{\` e}ncies del Cosmos (ICC), Universitat de Barcelona (IEEC-UB), Mart{\' i} i Franqu{\` e}s 1, 08028, Barcelona, Spain}
\altaffiltext{8}{Spitzer Science Center, California Institute of Technology, Pasadena, CA, USA}
\altaffiltext{9}{Universidad Diego Portales, Chile}
\altaffiltext{10}{Infrared Processing and Analysis Center, California Institute of Technology, Pasadena, CA, USA}
\altaffiltext{11}{Chalmers University of Technology, Department of Earth and Space Sciences, Onsala Space Observatory, 43992 Onsala, Sweden}
\altaffiltext{12}{National Radio Astronomy Observatory, Green Bank, WV, USA}
\altaffiltext{13}{National Optical Astronomy Observatory, 950 North Cherry Avenue, Tucson, AZ 85719, USA}
\altaffiltext{14}{The SETI Institute, 189 Bernardo Ave, Suite 100, Mountain View, CA 94043, USA}

\begin{abstract}
We present new IRAM 30m spectroscopic observations of the $\sim88$ GHz band, including emission from the \CCHn{1}{0} multiplet, \HCNj{1}{0}, \HCOj{1}{0}, and \HNCj{1}{0}, for a sample of 58 local luminous and ultraluminous infrared galaxies from the Great Observatories All-sky LIRG Survey (GOALS).
By combining our new IRAM data with literature data and Spitzer/IRS spectroscopy, we study the correspondence between these putative tracers of dense gas and the relative contribution of active galactic nuclei (AGN) and star formation to the mid-infrared luminosity of each system.
We find the \HCN{1}{0} emission to be enhanced in AGN-dominated systems ($\langle$~\Lprime{HCN}{1}{0}/\Lprime{HCO$^+$}{1}{0}$~\rangle~=1.84$), compared to composite and starburst-dominated systems ($\langle$~\Lprime{HCN}{1}{0}/\Lprime{HCO$^+$}{1}{0}$~\rangle~=1.14$, and 0.88, respectively).
However, some composite and starburst systems have \Lprime{HCN}{1}{0}/\Lprime{HCO$^+$}{1}{0} ratios comparable to those of AGN, indicating that enhanced HCN emission is not uniquely associated with energetically dominant AGN.
After removing AGN-dominated systems from the sample, we find a linear relationship (within the uncertainties) between $\log_{10}$(\Lprime{HCN}{1}{0}) and $\log_{10}$(\LIR), consistent with most previous findings.
\Lprime{HCN}{1}{0}/\LIR, typically interpreted as the dense gas depletion time, appears to have no systematic trend with \LIR for our sample of luminous and ultraluminous infrared galaxies, and has significant scatter.
The galaxy-integrated \HCN{1}{0} and \HCO{1}{0} emission do not appear to have a simple interpretation, in terms of the AGN dominance or the star formation rate, and are likely determined by multiple processes, including density and radiative effects.
\end{abstract}


\section{Introduction}
\label{sec:Introduction}

Molecular gas is observationally linked to ongoing star formation through observed correlations between the star formation rate surface density and the H$_2$ surface density as inferred from CO observations \citep[e.g.,][]{Bigiel2008,Leroy2012}.
\CO{1}{0} has a relatively low critical density (n$_{crit}\approx2\times10^3$ \cmc) and so traces the bulk of the molecular gas.
Molecular transitions such as \HCN{1}{0} and \HCO{1}{0} have critical densities n$_{crit}\approx3\times10^6$ \cmc and $2\times10^5$ \cmc, respectively, at 30 K, and so they are associated with higher density molecular hydrogen.
Early studies found a linear correlation between \HCN{1}{0} and the infrared luminosity---\LIRr---of galaxies \citep{Solomon1992,Gao2004a}.
This relation, which is tighter than that for \CO{1}{0} with \LIR, was interpreted as evidence that \HCN{1}{0} traces the dense gas directly associated with star formation.
Revisiting the relationship between SFR and this molecular tracer, \citet{GarciaBurillo2012} find \LFIR to be a super-linear function of \HCN{1}{0}, suggesting that there are other physical factors that are important for the HCN emission.

Calling into question the use of \HCN{1}{0} as a tracer of dense gas, several studies of systems hosting active galactic nuclei (AGN) have found integrated \citep{GraciaCarpio2006,Krips2008} and spatially-resolved enhancements \citep{Kohno2003,Imanishi2006,Imanishi2007,Imanishi2009,Davies2012} of \HCN{1}{0} emission compared to what is observed in starburst galaxies.
Similar results have been found for \HCN{4}{3} \citep{Imanishi2013,Imanishi2014a}.
These results suggest that HCN emission is enhanced in the presence of AGN, potentially invalidating the use of HCN as a tracer of the dense molecular gas associated with ongoing star formation, particularly in systems with AGN.
However it is notable that some systems with a known AGN do not show this enhanced ratio (ARP 299, \citealt{Imanishi2006a}; I Zw 1, \citealt{Evans2006}), suggesting the \HCN{1}{0} emission enhancement observed in some AGN hosts may have a more nuanced interpretation.

Modeling by various authors suggests the HCN and HCO$^+$ emission are affected by density, radiative, and abundance/ionization effects which potentially complicates interpretation of the line ratios in both photon dominated and X-ray dominated regions \citep[PDR and XDR, respectively; e.g.,][]{Aalto1994,Aalto1995,Huettemeister1995,Lepp1996,Meijerink2007}.
The possible influence of the XDR on molecular abundances has been used to argue that elevated \HCN{1}{0} is a signpost of an AGN, however \citet{Juneau2009} and \citet{Costagliola2011} argue the excitation of HCN and HCO$^+$ is not solely driven by abundance and so this ratio may not solely trace XDRs.
The relative influence of the other effects (infrared pumping, source compactness) has not been fully established.

In order to investigate the relation between the presence and strength of an AGN, and the HCN and HCO$^+$ emission in a large sample of gas-rich galaxies, we have observed 58 luminous and ultraluminous infrared galaxies ((U)LIRGs; \LIR $>10^{11}$ \Lsun), selected from the Great Observatories All-sky LIRG Survey \citep[GOALS;][]{Armus2009}, with the Institut de Radioastronomie Millim{\'e}trique (IRAM) 30m Eight Mixer Receiver \citep[EMIR;][]{Carter2012}.
These new observations (Section~\ref{sec:data}) of the molecular rotational transitions, \HCN{1}{0} and \HCO{1}{0}, are used in combination with calibrated AGN strengths determined from Spitzer/IRS \citep{Houck2004} spectroscopy of the GOALS sample \citep{Stierwalt2013,Inami2013} to assess the correlation of AGN strength with the global \HCN{1}{0} and \HCO{1}{0} emission (Section~\ref{sec:results}).
We then discuss the possible explanations for enhanced \HCN{1}{0} (Section~\ref{sec:discuss}).
Finally, we explore the utility of \HCN{1}{0} and \HCO{1}{0} as tracers of the mass of dense gas associated with star formation (Section~\ref{sec:sf}).

The power of this study comes from the increased sample size of (U)LIRGs with measurements of these lines and, particularly, from the large number of sources with measured mid-infrared diagnostic of the relative contributions of AGN and star formation to the infrared luminosity of each system \citep[a factor of 4 increase over][]{Costagliola2011}.
The use of the mid-infrared diagnostics facilitates a good estimation of the importance of an AGN to the mid-infrared emission (i.e., as opposed to simply using rudimentary optical ``AGN'' and ``starburst'' diagnostics to classify systems).
This enables a direct investigation of the global \Lprime{HCN}{1}{0}/\Lprime{HCO$^+$}{1}{0} line ratio (hereafter, HCN/HCO$^+$) as a function of the contribution of the AGN to the bolometric luminosity.

Throughout the paper we adopt a WMAP-5 cosmology \citep[H$_0$ = 70 \kms Mpc$^{-1}$, $\Omega_{\textnormal{vacuum}}=0.72$, $\Omega_{\textnormal{matter}}=0.28$]{Hinshaw2009}, with velocities corrected for the 3-attractor model of \citet{Mould2000}.

\section{Sample, Observations, and Data}
\label{sec:data}

\subsection{GOALS}

As noted, the data presented here were obtained as part of a millimeter survey of (U)LIRGs selected from GOALS.
The GOALS sample as a whole consists of all the (U)LIRGs from the Revised Bright Galaxy Sample \citep[RBGS; i.e., 60 $\mu m$ flux density greater than 5.24 Jy and \LIR $> 10^{11}$ \Lsun;][]{Sanders2003}.
GOALS\footnote{\url{http://goals.ipac.caltech.edu}} is a multi-wavelength survey aimed at understanding the physical conditions and activity in the most luminous galaxies in the local Universe.
The dataset includes spectroscopic and imaging observations in the infrared from Spitzer \citep[J. M. Mazzarella et al.\ \emph{in prep}]{Petric2011,Inami2013,Stierwalt2013} and Herschel \citep[][]{Diaz-Santos2013,Diaz-Santos2014,Lu2014}, GALEX and HST UV, optical, and near-infrared imaging \citep[A. S. Evans et al. \emph{in prep}]{Howell2010,Haan2011,Kim2013}, Chandra X-ray observations \citep{Iwasawa2011}, and a suite of ground-based radio and sub-millimeter observations \cite[e.g.,][Herrero-Illana et al. \emph{in prep}]{Leroy2011,Barcos-Munoz2015}.
These observations collectively trace the obscured and unobscured activity and constrain the structural properties and merger stages of these systems.

The objects in this study were selected from two portions of GOALS; a high-luminosity sample (\LIR $\geq 10^{11.4}$ \Lsun) to capture the most extreme star forming systems, and a sample of LIRGs with \LIR $\leq 10^{11.4}$ \Lsun which were selected to be isolated or non-interacting, on the basis of their stellar morphology \citep{Stierwalt2013}.
The combination of these subsets of GOALS ensures this sample spans the range of \LIR and merger stage within the GOALS sample as a whole.

\subsection{Observations and Data Reduction}

The IRAM 30m Telescope was used with the EMIR receiver to observe an 8 GHz instantaneous bandwidth, tuned to simultaneously capture \HCNj{1}{0}, \HCOj{1}{0}, the \CCHn{1}{0} multiplet, and \HNCj{1}{0}, thus reducing systematic uncertainties when comparing the fluxes of these lines.
The rest frequencies\footnote{obtained from the Splatalogue database: \url{http://splatalogue.net}} for these transitions are: $\nu_{rest}$(\HCN{1}{0}) $ = 88.631$ GHz, $\nu_{rest}$(\HCO{1}{0}) $ = 89.189$ GHz, and $\nu_{rest}$(\HNC{1}{0}) $ = 90.663$ GHz.
For CCH, we adopt a rest frequency which is the simple arithmetic mean of the individual multiplet frequencies, $\nu_{rest}$(CCH) $ = 87.370$ GHz.
The beam FWHM for these measurements is $\sim28"$, corresponding to a linear size of 16.8 kpc at the mean redshift of our sample -- thus we obtain galaxy-integrated measurements.
Observations were done in wobbler switching mode ($\sim1'$ throw, 0.8 s switching) to improve the baseline calibration.
In Table~\ref{table:obs} we list the pointing coordinates, assumed redshift, observing year \& month, integration times, system temperatures, and the backend used, for each source.

\begin{figure*}
\includegraphics[width=\textwidth]{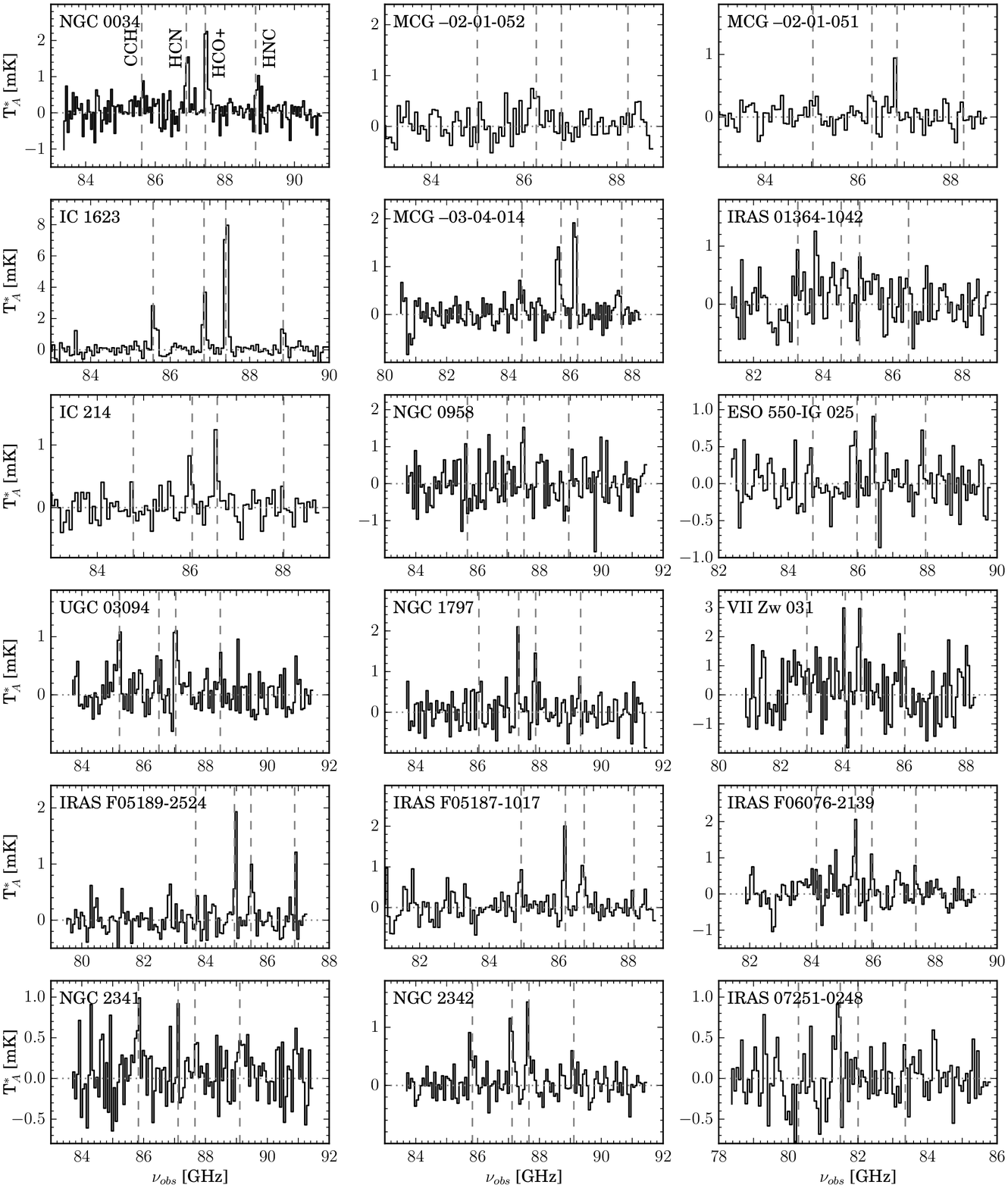}
\caption{IRAM 30m EMIR spectra of the systems observed as part of this program, sorted by Right Ascension.
In all panels we plot the observed T$_A^*$ as a function of the observed frequency.
The typical velocity resolution of these smoothed data is $60$ MHz, corresponding to roughly $215$ \kms.
Dashed vertical lines mark (from left to right) the expected locations of CCH, \HCN{1}{0}, \HCO{1}{0}, and \HNC{1}{0}, based on optical redshifts from NED, except for IRAS F16164-0746, where we have used our new measured redshift.
Information on the observing parameters are provided in Table~\ref{table:obs}.
Integrated fluxes (or upper limits) for each line are given in Table~\ref{table:fluxes}.
\emph{This figure is continued at the end of the manuscript, in the Figures currently labeled 8--10.}}
\label{fig:spectra}
\end{figure*}

Some of our observations use the FTS backend, which consists of 24 individual fourier transform spectrometer units.
In some scans, the FTS units exhibited gain variations, resulting in offsets of the baselines for individual units.
For those scans, linear fits were made to the baseline of each unit (masking out the expected locations of spectral lines), and the units were corrected to a common baseline.
Scans were combined and exported using the GILDAS/CLASS software\footnote{\url{http://www.iram.fr/IRAMFR/GILDAS}}.
Further analysis, including a linear baseline subtraction, smoothing, and flux measurement, was done using the \textit{pyspeckit} software\footnote{\url{http://pyspeckit.bitbucket.org}} \citep{Ginsburg2011c}.
The typical velocity resolution of these smoothed data is $60$ MHz, corresponding to roughly $215$ \kms.
In Figure~\ref{fig:spectra} we show the reduced spectra for all sources, with the expected positions of the CCH, \HCN{1}{0}, \HCO{1}{0}, and \HNC{1}{0} lines marked.
Lines were identified visually based on cataloged optical redshifts and total fluxes were determined by integrating under the line.
IRAS F16164-0746 had spectra which were clearly shifted from the location expected based on the cataloged redshift.
We measured a new redshift of $0.0229$ for IRAS F16164-0746 compared to the value from NED, $0.0272$.

We determined uncertainties in the fluxes by calculating the RMS, $\sigma_{ch}$, per channel of width $dv$, in the line-free regions of the spectrum and multiplying by the square root of the number of channels, $N_{ch}$ covered by each line.
We considered a line a detection if the measured flux exceeded $3\sigma_{ch} dv \sqrt{N_{ch}}$.
For non-detections we calculated a $3\sigma$ upper limit by assuming a square line profile with a width half that of the detected \HCN{1}{0} or \HCO{1}{0} line, or $200$ \kms if no lines were detected for that source.

For IRAM 30m observations, temperatures are reported on the T$^*_A$ scale; these values were converted to a main beam temperature, T$_{mb}=$T$^*_A/($B$_{eff}$/F$_{eff})$ using the main beam efficiency, B$_{eff}=0.81$, and a forward efficiency, F$_{eff}=0.95$, both measured by IRAM staff on 26 August 2013\footnote{\url{http://www.iram.es/IRAMES/mainWiki/Iram30mEfficiencies}}.
We use a Jy/K conversion of $3.906($F$_{eff}$/A$_{eff})=6.185$, where A$_{eff}=0.6$, to convert the reported T$^*_A$ values to flux densities.
Line luminosities were computed according to \citet[their Equation~3]{Solomon1992}, in units of K km s$^{-1}$ pc$^2$.

As noted, the wide EMIR bandwidth enables simultaneous measurements of \HCN{1}{0}, \HCO{1}{0}, \HNC{1}{0}, and CCH.
We had a detection rate of 78\%, 76\%, 35\%, and 37\%, for \HCN{1}{0}, \HCO{1}{0}, \HNC{1}{0}, and CCH, respectively. 
For detected lines the signal-to-noise weighted mean \HNC{1}{0}/\HCN{1}{0} and CCH/\HCN{1}{0} ratios are $0.5 \pm 0.3$ and $0.8 \pm 0.3$, respectively. 
One source, NGC 6285 has detected CCH emission and no detection of \HCN{1}{0} or \HCO{1}{0}; the CCH detection is $\sim3.8\sigma$ and, if confirmed, would be an interesting and rare example of a source with CCH brighter than HCN or HCO$^+$.
Such elevated CCH emission may indicate an overabundance of CCH or a lack of dense molecular gas \citep{Martin2014}.
The \HCN{1}{0}/\HCO{1}{0} ratio is discussed in more detail in Section \ref{sec:results}.
Integrated fluxes (or $3\sigma$ upper limits) are provided for all lines, in Table~\ref{table:fluxes}, but we limit the analysis here to the \HCN{1}{0} and \HCO{1}{0} lines.
Upper and lower limits in Figures use these $3\sigma$ limits.
\CO{1}{0} observations were also obtained as part of this program; these will be presented in R. Herrero-Illana et al.\ (\emph{in prep}).

In order to explore the influence of AGN on the \HCN{1}{0} and \HCO{1}{0} emission, we use the equivalent width (EQW) of the 6.2 $\mu m$ polycyclic aromatic hydrocarbon (PAH) measured from Spitzer/IRS low-resolution observations \citep{Stierwalt2013}.
The EQW of the $6.2~\mu m$ PAH feature compares the strength of the PAH emission associated with star formation with the strength of the mid-infrared continuum \citep[e.g.,][]{Genzel1998,Armus2004,Armus2007}.
The AGN is energetically more important in systems with lower values of the PAH EQW. 
Points in Figures \ref{fig:distance}--\ref{fig:sftracer} are color-coded by their $6.2~\mu m$ PAH EQW, to show the relationship between AGN and starburst dominated systems.

In addition to our new measurements, we incorporate observations from \citet{GraciaCarpio2006}, \citet{Costagliola2011}, and \citet{GarciaBurillo2012} in our analysis.
The combination of these three samples includes 63 individual galaxies with measured $6.2~\mu m$ PAH EQWs and detections in at least one of \HCN{1}{0} or \HCO{1}{0}, comprising roughly 25\% of the objects in the GOALS sample.

We use \LIR measurements from IRAS observations \citep{Armus2009}.
For galaxies in pairs, we assign a portion of \LIR to each component based on their $70$ or $24~\mu m$ flux ratio, as described by \citet{Diaz-Santos2013}.
Measurements of the \cii 158 $\mu m$ line are taken from \citet{Diaz-Santos2013}.

\subsubsection{Comparison with Previous Measurements}

Several of the sources observed by us have pre-existing published fluxes in the literature \citep[e.g.,][]{Gao2004,GraciaCarpio2006,GarciaBurillo2012}.
We compared our fluxes with those previous efforts and found good agreement for some sources (e.g., ARP 220, VV 114, NGC 0034, IRAS 15107+0724, IRAS 23365+3604), but we measure fluxes for some individual sources (NGC 2623, NGC 6701, UGC 5101, and NGC 7591) which are factors of $2-3$ higher than those previously published.
There appears to be no trend of these high fluxes with observing date or observing setup.
Despite some differences in fluxes for some sources, we find the HCN/HCO$^+$ ratios for those sources are in good agreement with previously published ratios.

\subsection{Distance and Aperture Effects}

\subsubsection{Source Distance Effects}

The $28''$ IRAM 30m beam FWHM covers linear scales of $6.1-46$ kpc for these sources.
Thus, these single-dish observations average together emission from many giant molecular clouds (GMCs) within each system, likely with varying physical conditions and environments.
This factor of $\sim8$ in distance leads to possible concerns about the effects of beam size on the results, particularly if there are systematic variations in the GMC properties or environments as a function of distance from the nuclei.

Previous interferometric observations with $\sim5''$ resolution have found the \HCN{1}{0} and \HCO{1}{0} emission to be unresolved in ULIRGs \citep{Imanishi2007,Imanishi2009}.
For the sources from their study that overlap with our sample, we find general agreement between the fluxes, suggesting the amount of extended flux is minimal and the HCN and HCO$^+$ emission is confined to a region smaller than the $28''$ beam of these IRAM 30m observations.

\begin{figure}
\includegraphics[width=0.5\textwidth]{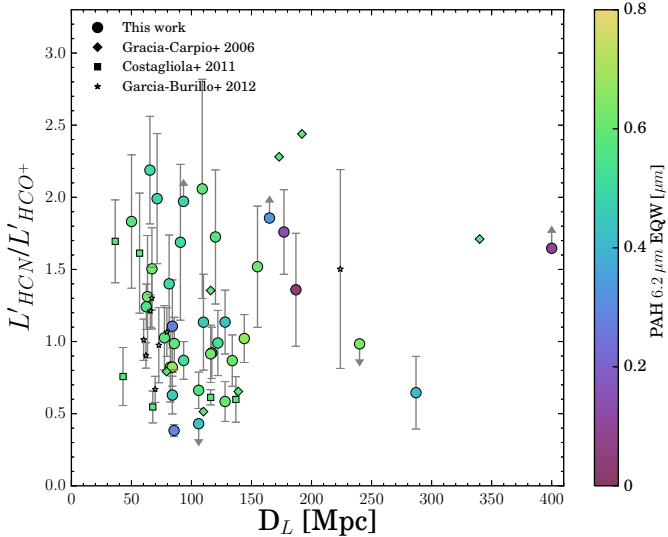}
\caption{Ratio of the \Lprime{HCN}{1}{0} and \Lprime{HCO$^+$}{1}{0} luminosities versus luminosity distance.
The lack of a significant correlation suggests our results are not being affected by the distances to each system.
Points are color-coded by the $6.2~\mu m$ PAH EQW.}
\label{fig:distance}
\end{figure}

In Figure~\ref{fig:distance} we compare the HCN/HCO$^+$ ratio with the luminosity distance; there is no obvious trend with distance, suggesting that the projected beam size is not the dominant factor in determining the HCN/HCO$^+$ ratio in our sample.

\subsubsection{Spitzer vs IRAM 30m Telescope Aperture Comparison}
\label{sec:aperture}

The $6.2~\mu m$ PAH EQW used to assess the relative dominance of the AGN in the mid-infrared was measured from a $3.6\arcsec$ aperture \citep{Stierwalt2013}.
Thus, these measurements are upper limits to the large-scale influence of the AGN, and we expect any systematic effect of this aperture difference between the mid-infrared and millimeter observations would serve to overestimate the importance of the AGN.

In other words, we would expect any signature of AGN-dominated gas to be diluted with increasing source distance. As will we shown (Section \ref{sec:results}), the excitation of these molecular transitions does not appear to be solely a function of the AGN strength, thus we conclude our results are not being significantly biased by the mis-match in aperture between the millimeter and mid-infrared datasets.

\section{Enhanced Global \HCN{1}{0} Emission does Not Uniquely Trace AGN Activity}
\label{sec:results}

\begin{figure}
\includegraphics[width=0.5\textwidth]{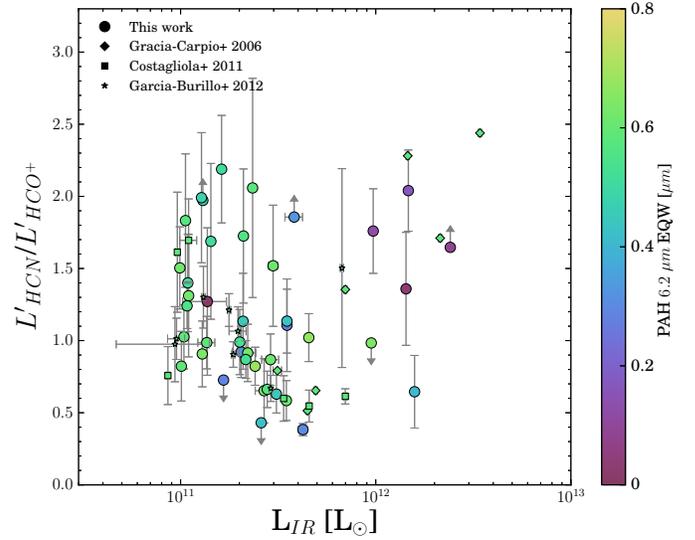}
\caption{Ratio of \Lprime{HCN}{1}{0} to \Lprime{HCO$^+$}{1}{0} luminosities versus the total infrared luminosity.
Points are color-coded by the $6.2~\mu m$ PAH EQW.}
\label{fig:LIR}
\end{figure}

Some of the initial claims that \HCN{1}{0} is enhanced in AGN were supported by plotting the \Lprime{HCN}{1}{0}/\Lprime{CO}{1}{0} and \Lprime{HCO$^+$}{1}{0}/\Lprime{CO}{1}{0} ratios as a function of \LIR.
In Figure~\ref{fig:LIR} we show the \HCN{1}{0}/\HCO{1}{0} ratio as a function of \LIR
\footnote{Here we do not compare with \CO{1}{0} as this would potentially include molecular gas which is not physically associated with the regions emitting in HCN (1--0) and HCO$^+$ (1--0).
For this study, we rely on the HCN/HCO$^+$ ratio alone, to avoid concerns with mis-matched apertures between the \CO{1}{0} measurements and the \HCN{1}{0} and \HCO{1}{0} (a 30\% difference in beam size).},
to compare we previous work.
As in previous studies \citep[e.g.,][]{GraciaCarpio2006}, we find a relative dearth of sources at high \LIR with low HCN/HCO$^+$.

However, our larger sample also contains a number of lower \LIR systems with HCN/HCO$^+$ ratios as high as those of ULIRGs ($> 1$), which suggests that increasing \LIR (or an associated increase in AGN contribution) may not be the sole driver of an increase in HCN/HCO$^+$ ratios.
A Spearman rank correlation analysis is consistent with the null hypothesis that HCN/HCO$^+$ and \LIR are uncorrelated ($\rho=0.021 \pm 0.178$, $z$-score$=0.057 \pm 0.494$; Table~\ref{table:spearman}).
The coefficients were computed using a Monte Carlo perturbation and bootstrapping method (with $10^5$ iterations), as described by \citet{Curran2014} and implemented in \citet{Curran2015}.

\begin{figure*}
\includegraphics[width=0.5\textwidth]{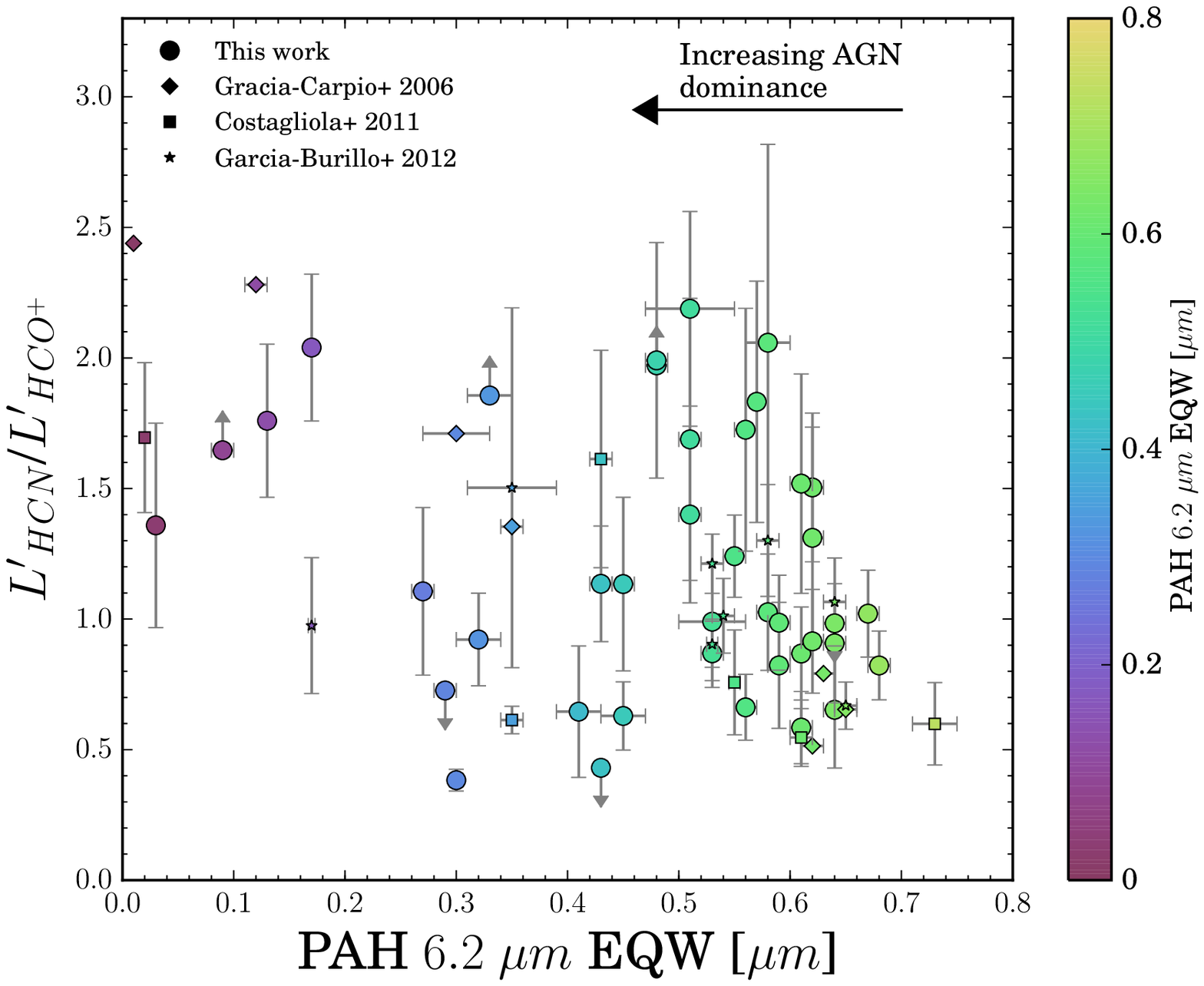}
\includegraphics[width=0.5\textwidth]{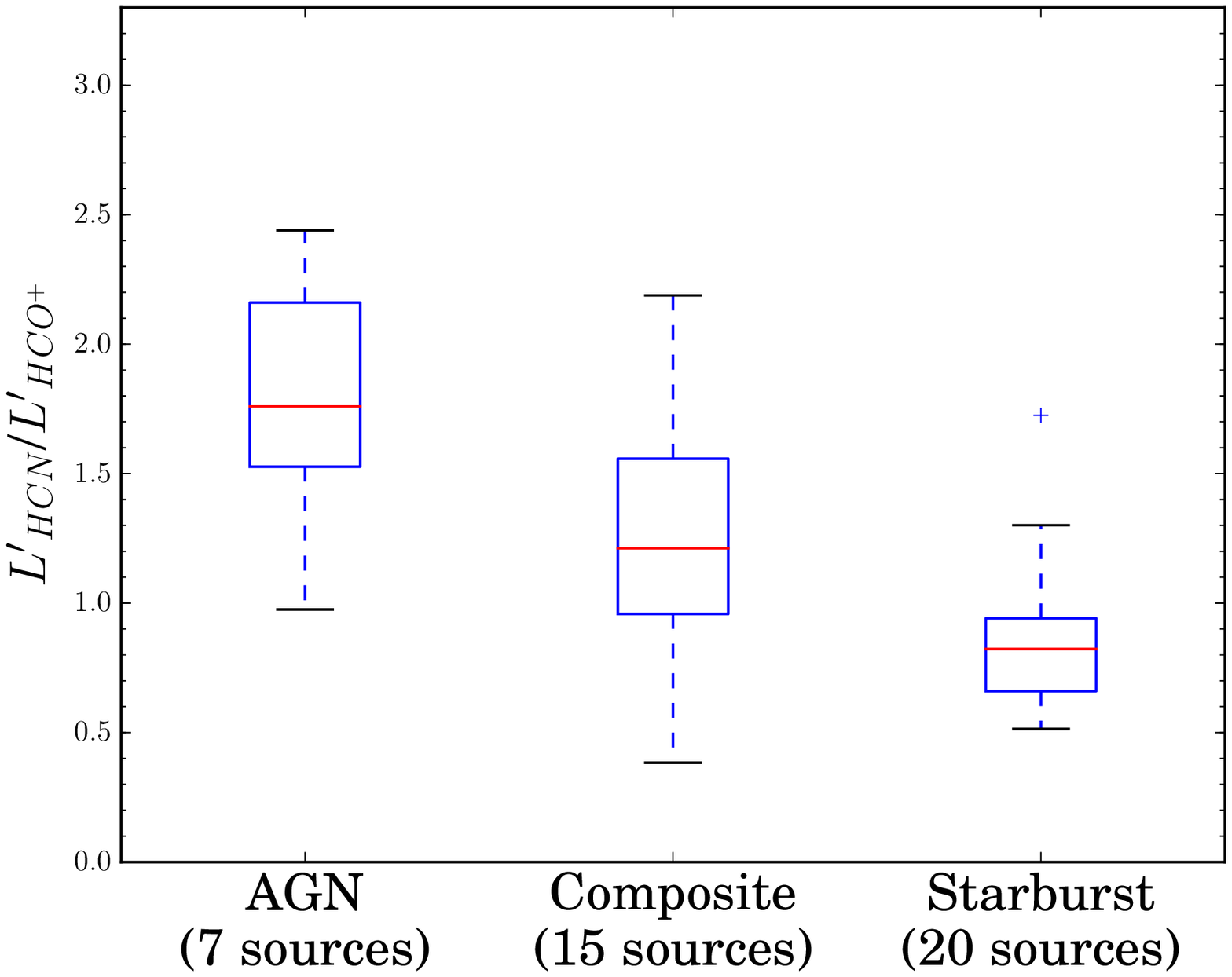}
\caption{Left: \Lprime{HCN}{1}{0}/\Lprime{HCO$^+$}{1}{0} as a function of the $6.2~\mu m$ PAH EQW from \citet{Stierwalt2013}.
The points are colored by their PAH EQW.
The symbol shape notes the origin of the millimeter line measurements; circles are new data (see Table~\ref{table:fluxes}), diamonds are from \citet{GraciaCarpio2006} and squares are from \citet{Costagliola2011}.
Right: Box plot showing the median (red line), interquartile range (boxes) and full range up to $1.5\times$ the interquartile range (IQR; black horizontal lines) for HCN/HCO$^+$ of the AGN dominated, composite, and starburst systems.
A flier (a point with a value $>1.5~\times$ IQR) is plotted with a '+' symbol.
Upper and lower limits were not included in the box plot.
HCN/HCO$^+$ is enhanced for systems which are AGN dominated in the mid-infrared, but some starburst dominated systems show similarly elevated ratios.
Weighted by signal-to-noise ratio, the average ratio for AGN dominated systems (PAH EQW $<~0.2~\mu m$), pure starbursts \citep[PAH EQW $> 0.55~\mu m$][]{Brandl2006}, and composite systems ($0.2~\mu m~<$ PAH EQW $<~0.55~\mu m$) is 1.84, 0.88, and 1.14, respectively.
}
\label{fig:AGNtracer}
\end{figure*}

The Spitzer/IRS observations of GOALS systems, presented by \citet{Stierwalt2013}, provide a calibrated measurement of the AGN contribution to the mid-infrared emission in each system, via the $6.2~\mu m$ PAH EQW.
In order to test claims that \HCN{1}{0} emission is enhanced in systems with AGN, we explore the HCN/HCO$^+$ ratio as a function of this AGN diagnostic.
In Figure~\ref{fig:AGNtracer}~(Left), we plot HCN/HCO$^+$ against the relative contribution of AGN and star formation to the infrared luminosity; where lower values of the PAH EQW indicate the presence of more energetically important AGN.
We consider systems with PAH EQW $<~0.2~\mu m$ to have energetically dominant AGN \citep[L$_{MIR,AGN}$/L$_{MIR} \gtrsim 60$ \%;) e.g.,][]{Petric2011}, while systems with PAH EQW $>~0.55~\mu m$ are considered starburst-dominated \citep{Brandl2006}, and systems with intermediate ratios as composites, where L$_{MIR}$ is the mid-infrared luminosity and L$_{MIR,AGN}$ is the AGN contribution to the mid-infrared luminosity.

We find that systems dominated by the AGN in the mid-infrared show elevated HCN/HCO$^+$ ratios, with a signal-to-noise ratio weighted mean ratio of $1.84\pm0.43$.
Starburst dominated systems have a weighted mean ratio of $0.88\pm0.28$, while composite systems have a weighted mean ratio of $1.14\pm0.49$.
However, for systems which appear to be star formation dominated, this ratio exhibits significant scatter.
Several starburst and composite systems have HCN/HCO$^+$ values which are comparable to the AGN dominated systems, suggesting that while energetically dominant AGN are associated with elevated \HCN{1}{0} emission alone, the converse is not true: enhanced \HCN{1}{0} emission does not imply the presence of an AGN.
We note that although these starburst and composite sources with enhanced \HCN{1}{0} emission have substantial uncertainties in their HCN/HCO$^+$ ratios, it is unlikely that the HCN/HCO$^+$ ratio is simultaneously overestimated for all six of these HCN-enhanced, composite/starburst sources.

A Spearman rank correlation analysis shows HCN/HCO$^+$ and the PAH EQW to be moderately anti-correlated, with $\rho=-0.512 \pm 0.127$ and a $z$-score of $-1.532\pm0.464$ (Table~\ref{table:spearman}).
In Figure~\ref{fig:AGNtracer}~(Right) we provide a ``box plot'' of the mean ratio, interquartile range, and full range of the HCN/HCO$^+$ values, separated by into source types based on the PAH EQW.
The distributions of HCN/HCO$^+$ for pure starbursts and AGN-dominated systems show a clear offset, with the AGN-dominated systems show a relative enhancement of \HCN{1}{0} emission over the pure starbursts.

It is worth noting there are fewer AGN-dominated sources than SB or composite sources; further observations of low PAH EQW systems would be useful to ensure the existing objects are representative.

\section{What Drives the Global \HCN{1}{0}/\HCO{1}{0} Ratio?}
\label{sec:discuss}

In Section~\ref{sec:results} we demonstrated that globally enhanced \HCN{1}{0} emission (relative to \HCO{1}{0}) is not correlated with the presence of an AGN (Figure~\ref{fig:AGNtracer}).
\citet{Costagliola2011} found similar results for a smaller sample of starbursts.
Is there a straight-forward explanation for the observed global \HCN{1}{0} emission in local (U)LIRGs? 
In the following subsections we explore a series of proposed explanations for enhanced HCN, including X-ray induced chemistry (Section \ref{sec:xrays}), the presence of a compact, high-density source (Section \ref{sec:compact}), and radiative pumping from absorption of mid-infrared photons (Section \ref{sec:pumping}).
We finish by discussing the possibility that the PAH EQW is not an ideal tracer of AGN (Section \ref{sec:AGNSB}) and then mention future observations which could be used to improve our understanding of HCN enhancements (Section \ref{sec:obs}).

\subsection{Comparison of \HCN{1}{0}/\HCO{1}{0} with X-ray Properties}
\label{sec:xrays}

To consider the possible influence of XDRs resulting from powerful AGN, we investigated the HCN/HCO$^+$ ratio as a function of X-ray properties, such as the hardness ratio and the total 0.5--10 keV X-ray luminosity, from Chandra X-ray observations of the GOALS sample \citep{Iwasawa2011}.
The HCN/HCO$^+$ ratio showed no correlation with either the hardness ratio or the X-ray luminosity, albeit with only ten sources common to both samples.
Additional X-ray observations would be useful to further investigate the influence of XDRs.
However, based on the currently available X-ray data, it does not appear that either the hardness of the X-ray spectrum or the total X-ray luminosity correlate with HCN/HCO$^+$, suggesting an XDR is not generally a major driver in enhancing \HCN{1}{0}, for these systems, possibly because the XDRs are spatially disconnected from the regions that dominate the global line luminosity.

A complication is that enhancement from X-rays may be most effective in obscured AGN and diagnosing this activity is difficult with observations below 10 keV.
Hard X-ray observations with NuSTAR \citep{Harrison2013}, though time-consuming, would provide an interesting test for the presence of obscured AGN in sources with enhanced HCN emission.

\subsection{HCN/HCO$^+$ Enhancements Through Source Compactness}
\label{sec:compact}

\begin{figure}
\includegraphics[width=0.5\textwidth]{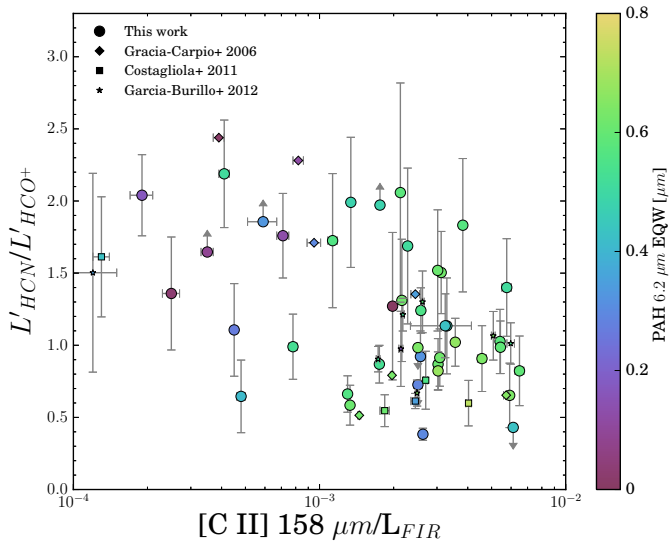}
\caption{A comparison of HCN/HCO$^+$ with \cii/\LFIR, as a proxy for the compactness of the source.
Systems with significant \cii deficits are located towards the left side of the plot.
There is a dearth of sources with low \cii/\LFIR values and low HCN/HCO$^+$ ratios, but the scatter is significant and some systems with high \cii/\LFIR values (less compact starbursts) still show elevated HCN/HCO$^+$ ratios, consistent with the mean of systems with substantial deficits (\cii/\LFIR $< 10^{-3}$). However there is significant scatter.
Points are color-coded by the $6.2~\mu m$  PAH EQW.}
\label{fig:deficit}
\end{figure}

In compact environments \HCO{1}{0} appears to be more susceptible to self-absorption than \HCN{1}{0} \citep{Aalto2015a}, which would lead to an increase in the observed HCN/HCO$^+$ ratio.
To test if elevated ratios are primarily associated with dense systems, we compare, in Figure~\ref{fig:deficit}, the HCN/HCO$^+$ ratio with the \cii/\LFIR ratios from the central Herschel spaxel ($9''.4\times9''.4$) as measured by \citet{Diaz-Santos2013}.
In local galaxies, the \cii/\LFIR value decreases with increasing \LFIR and dust temperature, T$_{dust}$.
That the ``\cii deficit'' tracks T$_{dust}$ suggests the deficit is the result of compact nuclear activity.
\citet{Diaz-Santos2013} find this trend of a lower \cii/\LFIR ratio with decreasing source size is present when considering only starbursts, suggesting that the \cii deficit is driven by the compactness of the starburst, rather than dust heated by an AGN.

Therefore, if the \HCN{1}{0}/\HCO{1}{0} ratio is being driven primarily by the density / compactness of the starburst, we should see a correlation between the two quantities.
For sources with \cii/\LFIR $>~10^{-3}$, the signal-to-noise weighted mean \Lprime{HCN}{1}{0} to \Lprime{HCO$^+$}{1}{0} is $1.0\pm0.4$, while sources with \cii/\LFIR $<~10^{-3}$ (large \cii deficits) have a mean ratio of $1.7\pm0.5$.
Formally, the ratio does not appear to vary as a function of \cii/\LFIR though a Spearman rank analysis suggests a moderate anti-correlation ($\rho=-0.401\pm0.142$, $z$-score$=-1.165 \pm 0.458$; Table~\ref{table:spearman}), though somewhat weaker than the anti-correlation of HCN/HCO$^+$ with PAH EQW.
The majority of sources with \cii/\LFIR $<~10^{-3}$ have HCN/HCO$^+$ ratios $ >1$, consistent with a scenario in which a compact and dense starburst causes an enhancement of \HCN{1}{0}.
On the other hand, a substantial number of systems without significant \cii/\LFIR deficits also have HCN/HCO$^+$ $> 1$.
For composite and starburst systems (PAH EQW $>~0.2~\mu m$) higher HCN/HCO$^+$ ratios do not appear to be associated with lower \cii/\LFIR values. Based on the available data, we cannot directly link the \cii deficit with enhanced \HCN{1}{0} emission.

It is worth noting that \HCN{1}{0} can be enhanced in sources that do not show a strong \cii deficit; if a compact continuum source is surrounded by extended disk-like star formation, the system may have a ``normal'' \cii/\LFIR ratio, but still show enhanced \HCN{1}{0} associated with the compact starburst on scales which cannot be resolved by Herschel.
See \citet{Diaz-Santos2014} for a discussion on extended \cii emission in the GOALS sample.
Assessing the spatial distribution of the \HCN{1}{0}/\HCO{1}{0} would be a useful comparison with \cii/\LFIR -- do regions with ``normal'' \cii/\LFIR correspond to regions with low \HCN{1}{0}/\HCO{1}{0}?

A possible concern with the above analysis is that the \LIR, \cii/\LFIR, and the PAH EQW are all correlated.
Indeed, the \LIR, the \cii deficit, and the $6.2~\mu m$ PAH EQW are interrelated, in that the most luminous sources tend to have the strongest \cii deficits \citep[][because both \LIR and \cii/\LFIR correlate with the dust temperature;]{Diaz-Santos2013} and lower PAH EQWs \citep[][]{Petric2011,Stierwalt2013}.
However, the converse is not true; a normal \cii/\LFIR value does not guarantee a high PAH EQW.
Systems with low or intermediate PAH EQWs are distributed across the range of \LIR and \cii/\LFIR values seen for our sample \citep[e.g., Figure~\ref{fig:deficit} and ][]{Diaz-Santos2013}..
In contrast, essentially all the high PAH EQW (starburst) systems have normal \cii/\LFIR and \LIR$\sim$ a few $\times 10^{11}$ \Lsun.
Thus, while these quantities are generally related, systems with low or intermediate PAH equivalent widths appear to have a range of compactness (as diagnosed by \cii/\LFIR) and so the PAH EQW and \cii deficit appear to be semi-independent.
That the \HCN{1}{0}/\HCO{1}{0} ratio does not correlate well with either, suggests the origin of HCN enhancement cannot be easily assigned solely to AGN or to the presence of a compact starburst.

\subsection{HCN/HCO$^+$ Enhancements Through Mid-infrared Pumping}
\label{sec:pumping}

The strong mid-infrared continuum present in many (U)LIRGs may also influence these line ratios.
The HCN molecule has degenerate bending modes in the infrared and can absorb mid-infrared photons ($14~\mu m$) to its first vibrational state.
The transitions have high level energies ($>1000$ K) and mid-IR emission with a brightness temperature of at least 100 K is necessary to excite them \citep{Aalto2007}.
A sign that infrared pumping of HCN is taking place is the presence of line emission from rotational transitions within the vibrational band \citep[e.g.,][]{Sakamoto2010}.
However, it may be possible for HCN to be infrared pumped without detectable emission from these rotational-vibrational lines.
When the brightness temperature is close to the lower limit for pumping, the excitation of the rotational-vibrational line is so low that it takes a very large column density to result in a large enough optical depth for this rotational-vibrational line to be detected.

Several sources in our sample do show evidence for infrared absorption at $14~\mu m$ which could be associated with pumping of HCN \citep{Lahuis2007,Sakamoto2010}, but none of the starburst systems with enhanced HCN emission show $14~\mu m$ absorption in the Spitzer/IRS observations from \citet{Inami2013}.

HCO$^+$ is similarly susceptible to infrared pumping, via a $\sim12~\mu m$ line; using the IRS high-resolution spectroscopy from \citet{Inami2013} we searched for a $\sim12~\mu m$ absorption feature in those systems with high ratios, where we postulated mid-infrared pumping could be important.
We did not find any evidence for HCO$^+$ absorption.

However, it is not clear that existing mid-infrared observations are sensitive enough to establish firm upper limits on the importance of infrared pumping.
Existing spectroscopy may not have the sensitivity to identify absorption which is sufficient to affect level populations.

\subsection{Is the PAH EQW A Good Tracer of AGN Strength}
\label{sec:AGNSB}

An obvious question is whether or not the PAH EQW is a robust measure of AGN strength in (U)LIRGs.
\citet{Aalto2015a} present interferometric observations of several objects, including the composite source CGCG  049-057, one of our sources with enhanced HCN emission.
The \HCN{3}{2} $v_2=0$ and \HCO{3}{2} lines show evidence for significant self absorption and the \HCN{3}{2} $v_2=1$ ro-vibrational line is detected (consistent with mid-infrared pumping).
\citet{Aalto2015a} interpret this as evidence for a very compact (tens of pc) warm ($> 100$ K) region surrounded by a large envelope of cooler, more diffuse gas.
The total column density through these systems is likely quite high ($> 10^{24}$ \cmcol).
If the column densities are sufficiently high to support mid-infrared pumping, the optical depths in the infrared may be high enough that the mid-infrared continuum level does not trace the intrinsic energy source -- thus the PAH EQW diagnostic may miss highly embedded AGN.

In Section \ref{sec:aperture} we note that the PAH EQW is an upper limit to the AGN's influence on large scales, but this measurement could also be considered as a lower limit to the AGN's influence on smaller scales.
Thus, consistent with results of \citet{Aalto2015a} for CGCG  047-059, it is possible that these starburst or composite sources with high \HCN{1}{0}/\HCO{1}{0} ratios host embedded AGN, and substantial mid-infrared optical depths result in an under-estimate of the AGN's influence when using the PAH EQW.

\subsection{Future Observations}
\label{sec:obs}

Though we have shown in this study that some starburst-dominated systems have global \Lprime{HCN}{1}{0}/\Lprime{HCO$^+$}{1}{0} ratios in excess of unity, the starburst-dominated systems outnumber AGN-dominated systems in the present sample.
More IRAM 30m single-dish observations of AGN-dominated (U)LIRGs would be useful to study the dense-gas tracers in that population, particularly to see if the absence of AGN-dominated sources with low \Lprime{HCN}{1}{0}/\Lprime{HCO$^+$}{1}{0} ratios is truly representative or merely reflects small number statistics.

Spatially resolved comparisons of the HCN emission with CO or HCO$^+$ have been undertaken for systems known to host an AGN \citep[e.g.,][]{Kohno2003,Imanishi2006,Imanishi2007,Imanishi2009,Davies2012,Imanishi2013,Imanishi2014a}; it would be instructive to perform the same exercise on starburst dominated systems.
The Atacama Large Millimeter/sub-millimeter Array is the natural instrument for this, and such a study could investigate the spatial variation in the \Lprime{HCN}{1}{0}/\Lprime{HCO$^+$}{1}{0} ratio for starbursts.
If the ratio peaks on the nucleus but is low elsewhere (as in systems with AGN), does that correspond to a compact mid-infrared emitting region? Spatially resolving this ratio in starburst systems with both high and low ratios may enable a separation of the relative importance of source compactness and/or mid-infrared pumping, in setting the ratio.

\citet{Krips2008} studied the emission of multiple HCN and HCO$^+$ transitions for twelve systems, finding evidence for systematic differences in the \HCN{1}{0}/\HCO{1}{0} and \HCN{3}{2}/\HCO{3}{2} ratios between AGN and starbursts, with AGN having values $>2$ for both ratios.
They also found evidence that the (3--2)/(1--0) ratio for both HCN and HCO$^+$ varies between AGN and starbursts.
The number of GOALS objects with (3--2) measurements of HCN and HCO$^+$ is currently too small to draw conclusions regarding the (3--2)/(1--0) ratios as a function of AGN strength, but it would be possibly illuminating to compare the higher J$_{upper}$ transitions with the AGN diagnostic used here.

The discovery of mid-infrared pumping and associated high-column densities in enhanced HCN sources suggests that even mid-infrared diagnostics such as the PAH EQW may miss highly embedded AGN.
If this is the case, hard X-ray observations may provide the only conclusive evidence for AGN in sources with heavily obscured nuclei.
However, the weakness of the $>10$ keV X-rays in ULIRG AGN \citep[e.g., Mrk 231;][]{Teng2014} may make these observations difficult with existing facilities.

Finally, the influence of infrared pumping is still uncertain.
Future mid-infrared observations of these sources with The James Webb Space Telescope would provide tighter constraints on the importance and ubiquity of mid-infrared pumping, for both HCN and HCO$^+$.
See also \citet{Aalto2007} for a discussion on using resolved observations to distinguish mid-infrared pumping and XDR scenarios.

\section{The Relationships between Molecular Line Luminosities, Star Formation Rates, and Gas Depletion Timescales}
\label{sec:sf}

\subsection{Star Formation Rates}

\begin{figure*}
\includegraphics[width=0.5\textwidth]{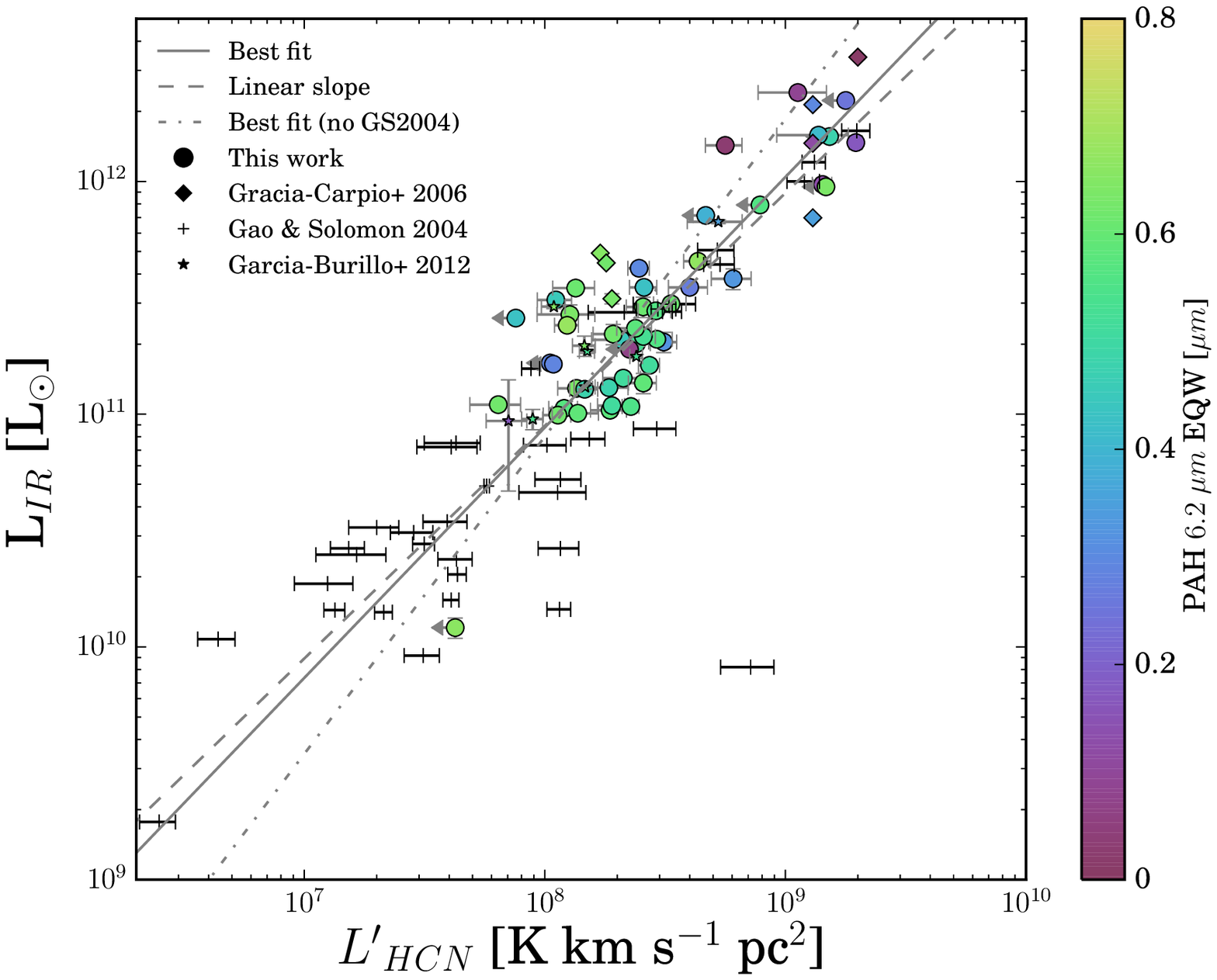}
\includegraphics[width=0.5\textwidth]{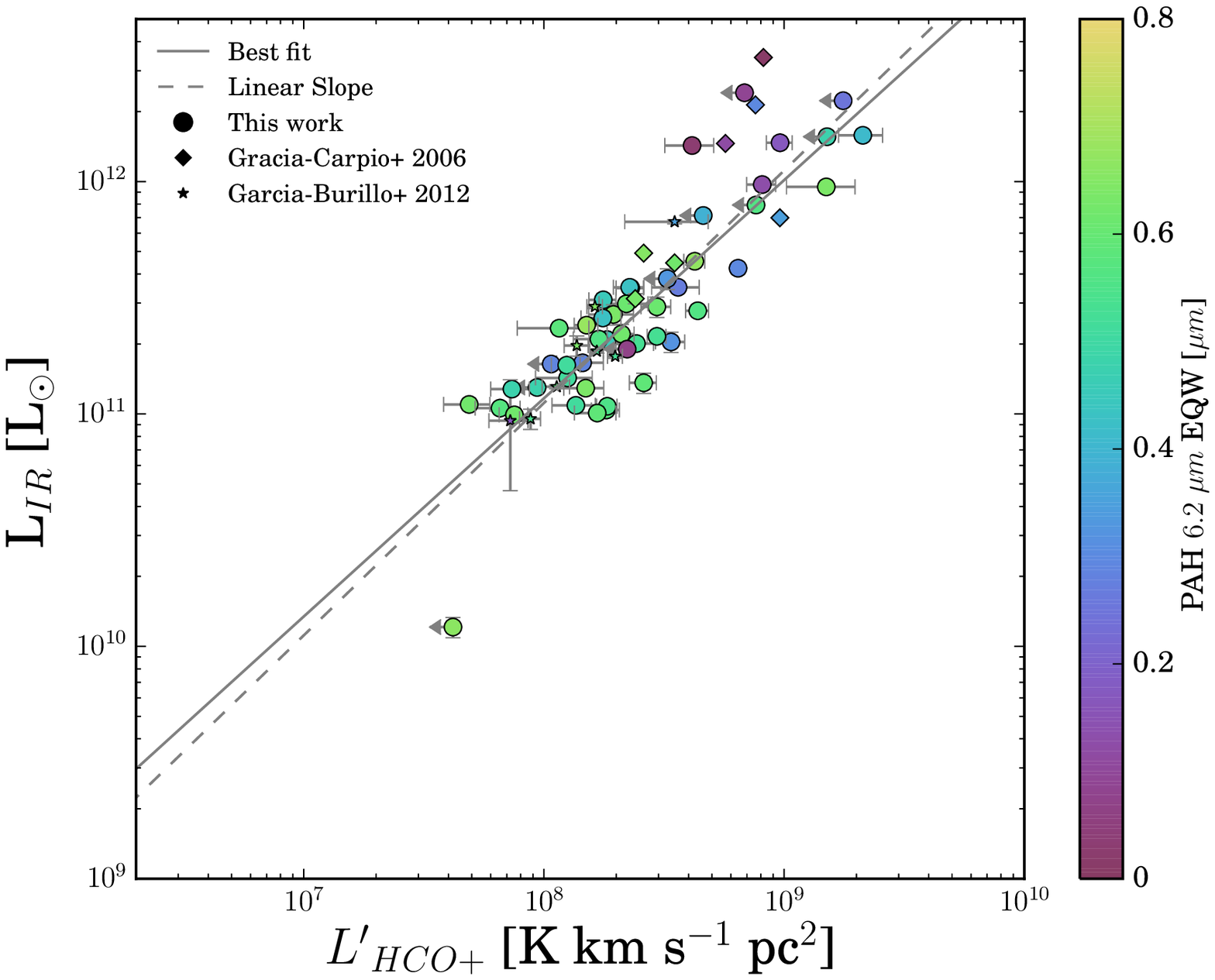}
\caption{\LIR plotted as a function of \Lprime{HCN}{1}{0} (Left) and \Lprime{HCO$^+$}{1}{0} (right).
The points are color-coded by 6.2 $\mu m$ EQW.
The solid lines in the left and right panels denote the best fit relations of equations~\ref{eq:sfrelation} and~\ref{eq:hcorelation}, respectively, while the dashed lines show the relation if we fix the slope to be linear.
The fits do not include upper limits, but the (solid and dashed) fits to the \LIR--\Lprime{HCN}{1}{0} relation (left) includes the data from \citet{Gao2004a}, corrected to our cosmology.
In the left panel, we show a fit excluding the \citet{Gao2004a} points, as the dot-dash line.
We find a slope consistent with being linear, when considering \LIR(\Lprime{HCN}{1}{0}) (Equation~\ref{eq:sfrelation}), consistent with \citet{Gao2004a}.
For \LIR(\Lprime{HCO$^+$}{1}{0}) we find a slope consistent with a linear relation (Equation~\ref{eq:hcorelation}).
The fits do not include upper limits or AGN-dominated systems (PAH EQW $<~0.2~\mu m$).}
\label{fig:SFrelation}
\end{figure*}

We now turn to the relationship between the \HCN{1}{0} emission and the star formation rates in (U)LIRGs.
While \HCN{1}{0} is generally taken to trace the mass of dense molecular gas, M$_{dense}$, previous studies have found conflicting results for the relationship between \LIR and \Lprime{HCN}{1}{0}.
In Figure~\ref{fig:SFrelation}~(Left) we plot the standard relation between \Lprime{HCN}{1}{0} and \LIR including additional measurements from \citet[with values corrected to our assumed cosmology and for the 3-attractor model of \citet{Mould2000}]{Gao2004a}.

We fit the relationship using a maximum likelihood technique, considering the errors in both \LIR and \Lprime{HCN}{1}{0} \citep[e.g., Sec 7 of][]{Hogg2010}.
When uncertainties for mm-line fluxes were not quoted in the literature, we assume 20\%.
Uncertainties on the fit parameters were determined using Markov Chain Monte Carlo (MCMC) sampling\footnote{Calculated using the \texttt{emcee} python library \citep{Foreman-Mackey2013}.} and quoted fit parameter uncertainties are for the 99\% confidence interval.

To this point, we have not considered systematic uncertainties in our new observations, as the simultaneous measurement of HCN and HCO$^+$ should result in significantly reduced systematic uncertainties in the HCN/HCO$^+$ ratio,in the HCN/HCO$^+$ ratio, compared to non-simultaneous measurements.
However, for the present comparison of \LIR with these lines, the absolute flux of these lines may be subject to systematic uncertainties.
We tested two approaches for dealing with systematic uncertainties: 1) assigning additional systematic uncertainties (equal in magnitude to the statistical uncertainties) to each datapoint\footnote{The values quoted by \citet{GarciaBurillo2012} already include systematic uncertainties, so we did not increase the uncertainties on those points.} and fitting the relationship, or 2) including only statistical uncertainties\footnote{In this case, we reduced the uncertainties on the \citet{GarciaBurillo2012} points to remove their estimate for the systematic uncertainties.} and allowing for additional uncertainties in the fitting process.
In addition to fitting for the slope and intercept, we included an additional ``nuisance'' parameter, $f$, in the likelihood function to fit for the fractional amount by which the uncertainties are underestimated (e.g., due to not explicitly including systematic uncertainties).
After the MCMC sampling, we marginalized over $f$ when determining the final uncertainties for the slope and intercept.
Both approaches yielded the same results for the best-fit relations, so we conclude that our fitting process appropriately accounts for non-statistical uncertainties and present the numerical results from the latter approach.
For consistency, we show only the statistical uncertainties for values plotted in Figures~\ref{fig:SFrelation} and \ref{fig:sftracer}.

Our best fit relation to our new data combined with the literature data (omitting both AGN-dominated systems, in which \LIR may not solely trace star formation, and \HCN{1}{0} upper limits) between \LIR and \Lprime{HCN}{1}{0} is:

\begin{equation}
\log_{10} \text{L}_{\text{IR}} = (1.08^{+0.18}_{-0.16}) \log_{10} \text{L}^{\prime}_{\text{HCN} (1-0)} + (2.32^{+1.54}_{-1.50})
\label{eq:sfrelation}
\end{equation}

This fit is shown in Figure~\ref{fig:SFrelation}~(Left) as the solid line.
We also fit \Lprime{HCN}{1}{0}--\LIR, fixing the slope to be linear, which we show in Figure~\ref{fig:SFrelation}~(Left) as a dashed line.
The slope of the \Lprime{HCN}{1}{0}--\LIR relation is consistent with being linear at the $\sim1.3\sigma$ level.
We note that \citet{GarciaBurillo2012} found the \Lprime{HCN}{1}{0}--\LFIRr relation to be steeper than linear, with a slope of $1.23\pm0.05$ (68 \% confidence interval).

In Figure~\ref{fig:SFrelation}~(Right) we perform the same comparison, but utilizing \Lprime{HCO$^+$}{1}{0} instead of \Lprime{HCN}{1}{0}.
As \HCO{1}{0} has a higher critical density than \CO{1}{0}, but lower than \HCN{1}{0}, it is useful to test if it is also linearly tracked by the SFR.
The \Lprime{HCO$^+$}{1}{0}--\LIR relation is also consistent with linear:

\begin{equation}
\log_{10} \text{L}_{\text{IR}} = (0.94^{+0.25}_{-0.19}) \log_{10} \text{L}^{\prime}_{\text{HCO}^+ (1-0)} + (3.54^{+2.18}_{-2.20})
\label{eq:hcorelation}
\end{equation}

Thus, our results are consistent with scenarios in which the HCN and HCO$^+$ emission both trace the dense gas associated with ongoing star formation, albeit with scatter, likely the result of one or more of the excitation mechanisms discussed in Section \ref{sec:discuss}.

\subsection{The Lack of HCO$^+$ Measurements for non-(U)LIRGs}

The inclusion of the \citet{Gao2004} HCN observations adds in a substantial number of galaxies with \LIR $< 10^{11}$ \Lsun, but corresponding \HCO{1}{0} observations are not available.
In order to assess the degree to which these lower-luminosity systems influence our fitting results, we also fit the \HCN{1}{0}--\LIR relation without the points from \citet{Gao2004}; this fit is shown as the dot-dash line in Fig \ref{fig:sftracer}~(Left) and the best-fit relation is:

\begin{equation}
\log_{10} \text{L}_{\text{IR}} = (1.09^{+0.48}_{-0.31}) \log_{10} \text{L}^{\prime}_{\text{HCN} (1-0)} + (2.30^{+3.37}_{-3.64})
\label{eq:hcnrelation-nogs}
\end{equation}

which is consistent with that for the full dataset.

\subsection{Dense Gas Depletion Times}

\begin{figure*}
\includegraphics[width=0.5\textwidth]{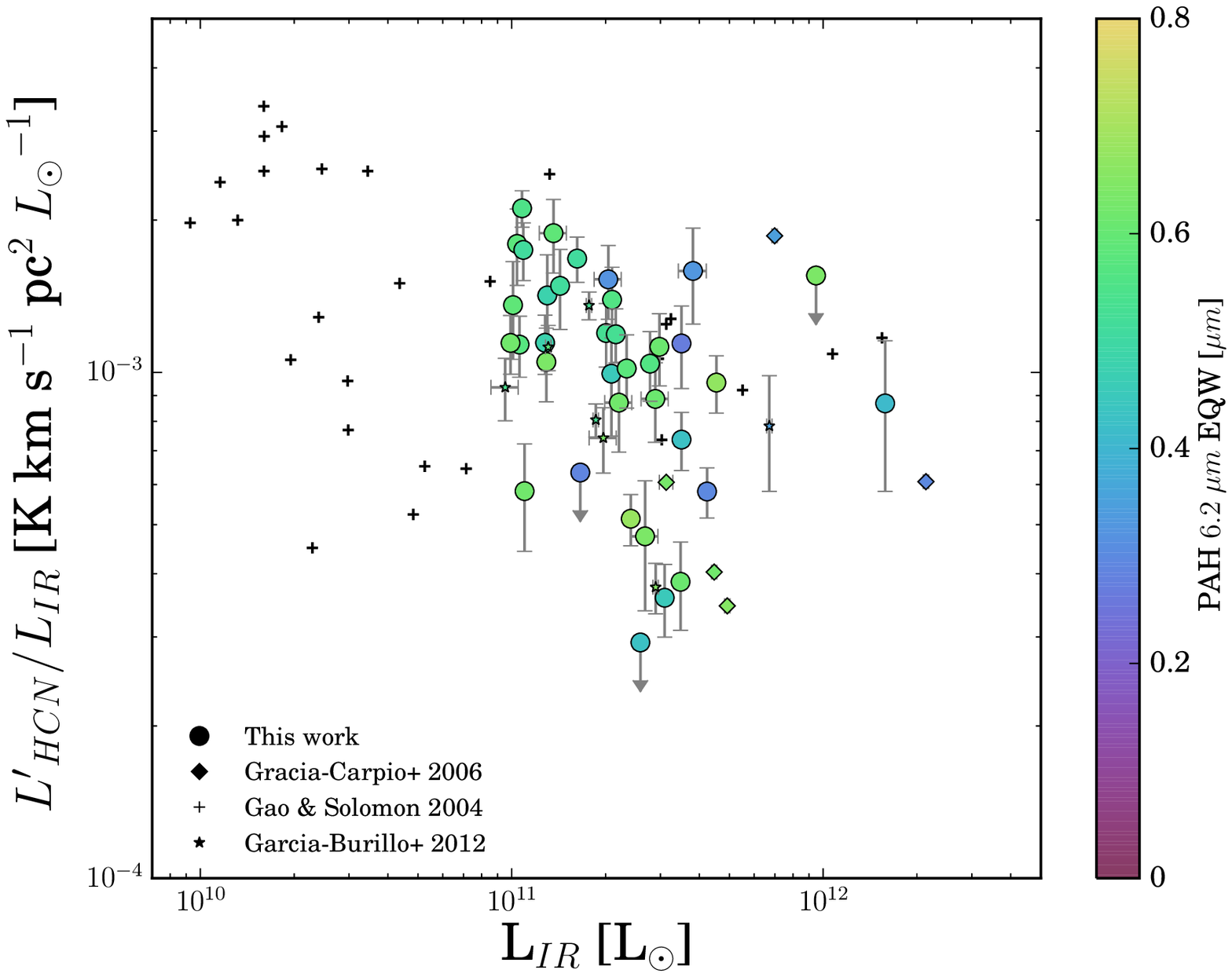}
\includegraphics[width=0.5\textwidth]{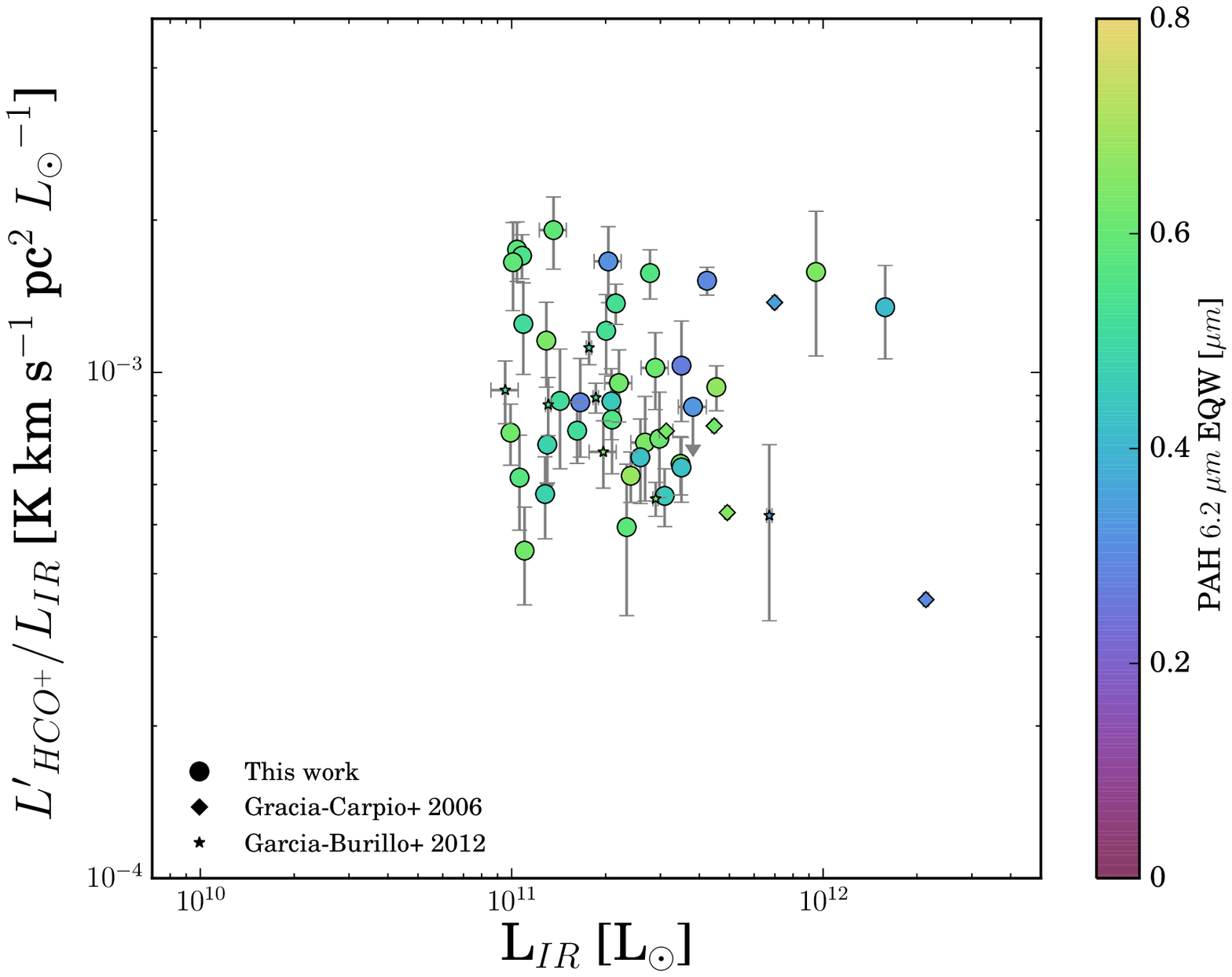}
\caption{Left: \Lprime{HCN}{1}{0}/\LIR versus \LIR.
The \citet{Gao2004} points have been corrected to the cosmology assumed here.
Right: \Lprime{HCO$^+$}{1}{0}/\LIR versus \LIR.
Points are color-coded by the $6.2~\mu m$ PAH EQW.
We see no clear trend of the gas depletion time with \LIR.}
\label{fig:sftracer}
\end{figure*}

The ratio \Lprime{HCN}{1}{0}/\LIR is often taken as $\propto$ M$_{\textnormal{\scriptsize{dense}}}$/SFR, which corresponds to the depletion time ($\tau_{dep}$) of the dense molecular gas.
In Figure~\ref{fig:sftracer}~(Left) we show \Lprime{HCN}{1}{0}/\LIR as a function of \LIR, excluding sources which are AGN dominated (PAH EQW $<~0.2~\mu m$).
As in Figure~\ref{fig:SFrelation}~(Left), we include observations from \citet{Gao2004}.
The mean $log_{10}($\Lprime{HCN}{1}{0}/\LIR$)=1\times10^{-3}$ with a RMS scatter of 0.22 dex.
The AGN and starburst dominated systems (as traced by the 6.2 $\mu m$ PAH feature) have similar \Lprime{HCN}{1}{0}/\LIR ratios.

Similarly, \Lprime{HCO$^+$}{1}{0}/\LIR shows no obvious trend with \LIR, but has substantial scatter.
The mean $log_{10}($\Lprime{HCO$^+$}{1}{0}/\LIR$)=1\times10^{-3}$ with a RMS scatter of 0.19 dex.

The significant scatter in \Lprime{HCN}{1}{0}/\Lprime{HCO$^+$}{1}{0} can plausibly be attributed to the influence of multiple excitation mechanisms for these dense gas tracers, as described above.
These observations are consistent with a scenario in which molecular abundance \citep{Lepp1996}, density \citep{Meijerink2007}, and radiative effects \citep{Aalto1995} all influence the global \HCN{1}{0} emission, obscuring a simple link between \Lprime{HCN}{1}{0} and the mass of dense gas directly associated with ongoing star formation.
Stated differently, the conversion factor between \Lprime{HCN}{1}{0} (or \Lprime{HCO$^+$}{1}{0}) and M$_{dense}$ likely depends on the relative HCN--H$_2$ abundance in addition to the overall density, excitation source (PDR vs XDR), and influence of infrared pumping.
A better understanding of the processes in (U)LIRGs contributing to \Lprime{HCN}{1}{0} and \Lprime{HCO$^+$}{1}{0} are needed to determine if the scatter in Figure~\ref{fig:sftracer} is due to differences in the consumption rates of molecular gas or a varying \Lprime{HCN}{1}{0}--M$_{dense}$ conversion factor.
Future observations (Section~\ref{sec:obs}), coupled with improved modeling, should be able to discriminate between these scenarios for the HCN and HCO$^+$ emission in (U)LIRGs.

\section{Summary}

We make use of new measurements of the putative high density gas tracers \HCN{1}{0} and \HCO{1}{0} in a sample of local (U)LIRGs. A comparison between the ratios of these lines and the $6.2~\mu m$ PAH EQW mid-infrared AGN indicator suggests enhancements in the global \HCN{1}{0} emission (relative to \HCO{1}{0}) does not uniquely trace the presence of an energetically dominant AGN.
While we find enhanced \HCN{1}{0} emission relative to \HCO{1}{0} in objects hosting dominant AGN, we find the same magnitude of enhancement is also possible for systems which are dominated by star formation.
The \HCN{1}{0} and \HCO{1}{0} emission does not seem to be driven by a single process.
It is likely that their emission is determined by the interplay of radiation field, gas column, and gas density.
This hampers a simple interpretation of the line ratio.
Existing data on the X-ray and mid-infrared properties of these systems are not complete or deep enough (respectively) for us to prefer one of XDRs or mid-infrared pumping as the mechanism for enhancing the \HCN{1}{0} emission.

We compare the \HCN{1}{0} emission with the star formation rate (\LIR) and find a linear relationship, consistent with some previous studies \citep[e.g.,][]{Solomon1992,Gao2004a}.
However, our result is also consistent with a superlinear relationship, within our 99\% confidence interval, analogous to the recent study utilizing \LFIR, by \citet{GarciaBurillo2012}.
The large scatter in the L$^{\prime}$/\LIR ratios is consistent with a scenario in which these dense gas tracers can be influenced by density effects, infrared pumping, and/or XDRs. This potentially complicates the determination of global dense gas masses and dense gas depletion times.

\acknowledgements

We thank the anonymous referee for their careful reading of our manuscript and for their helpful comments and suggestions, which improved the paper.

G.C.P. and A.S.E. were supported by the NSF grant AST 1109475, and by NASA through grants HST-GO10592.01-A and HST-GO11196.01-A from the Space Telescope Science Institute, which is operated by the Association of Universities for Research in Astronomy, Inc., under NASA contract NAS5-26555.
G.C.P. and E.T. were supported by the CONICYT Anillo project ACT1101 (EMBIGGEN).
G.C.P. was also supported by a Visiting Graduate Research Fellowship at the Infrared Processing and Analysis Center / Caltech and by a FONDECYT Postdoctoral Fellowship (No.\ 3150361).
This work was supported in part by National Science Foundation Grant No. PHYS-1066293 and the hospitality of the Aspen Center for Physics.
G.C.P. acknowledges the hospitality of the National Socio-environmental Synthesis Center (SESYNC), where portions of this manuscript were written.
A.S.E. was also supported by the Taiwan, R.O.C. Ministry of Science and technology grant MoST 102-2119-M-001-MY3.
KI acknowledges support by the Spanish MINECO under grant AYA2013-47447-C3-2-P and MDM-2014-0369 of ICCUB (Unidad de Excelencia ``Mar\'ia de Maeztu'').
RHI, MAPT, and AA also acknowledge support from the Spanish MINECO through grant AYA2012-38491-C02-02.
E.T. was also supported by the Center of Excellence in Astrophysics and Associated Technologies (PFB 06) and by the FONDECYT regular grant 1120061.

This research has made use of the NASA/IPAC Extragalactic Database (NED) which is operated by the Jet Propulsion Laboratory, California Institute of Technology, under contract with the National Aeronautics and Space Administration.
This research also made use of Astropy, a community-developed core Python package for Astronomy \citep{Astropy2013}, the \texttt{cubehelix} python library\footnote{\url{https://github.com/jradavenport/cubehelix}}, and NASA's Astrophysics Data System.
The Spitzer Space Telescope is operated by the Jet Propulsion Laboratory, California Institute of Technology, under NASA contract 1407.
G.C.P. thanks Ina Evans for a critique of the comments of an earlier version of this manuscript.

{\it Facility:} \facility{IRAM:30m (EMIR, WIMA, FTS)}

\begin{deluxetable}{lllccccl}
\tablecaption{IRAM 30m Observing Log}
\tablehead{\colhead{Source} &
 \colhead{J2000 RA} &
 \colhead{J2000 Dec} &
 \colhead{$z$} &
 \colhead{Obs Month} &
 \colhead{$t_{int}$} &
 \colhead{T$_{sys}$} &
 \colhead{Backend} \\
 &
 \colhead{[deg]} &
 \colhead{[deg]} &
 &
 \colhead{[YYYY-MM]} &
 \colhead{[min]} &
 \colhead{[K]} & }
\startdata
NGC 0034    & \phn\phn2.77729   & \phantom{-}--12.10775 & $0.0200$  & 2011-09   & $276$ & $116$  & WILMA \\
MCG --02-01-052 & \phn\phn4.70904   & \phantom{-}--10.36246 & $0.0274$  & 2011-12   & $127$ & \phn$89$   & FTS \\
MCG --02-01-051 & \phn\phn4.71208   & \phantom{-}--10.37672 & $0.0270$  & 2011-12   & $277$ & \phn$90$   & FTS \\
IC 1623 & \phn16.94804  & \phantom{-}--17.50721 & $0.0205$  & 2011-12   & \phn$95$  & \phn$88$   & FTS \\
MCG --03-04-014 & \phn17.53733  & \phantom{-}--16.85272 & $0.0342$  & 2011-12   & $223$ & \phn$86$   & FTS \\
IRAS 01364-1042 & \phn24.72050  & \phantom{-}--10.45317 & $0.0487$  & 2011-09   & $117$ & \phn$98$   & WILMA \\
IC 214  & \phn33.52279  & \phn+5.17367  & $0.0300$  & 2011-12   & $244$& $84$    & FTS \\
NGC 0958    & \phn37.67858  & \phn\phantom{-}--2.93905  & $0.0193$  & 2012-10   & \phn$61$  & $139$  & FTS \\
ESO 550-IG 025    & \phn65.33333  & \phantom{-}--18.81094 & $0.0321$  & 2013-08   & $298$ & $105$  & WILMA \\
UGC 03094   & \phn68.89096  & +19.17172 & $0.0247$  & 2012-10   & $224$ & $126$  & FTS \\
NGC 1797    & \phn76.93690  & \phn\phantom{-}--8.01911  & $0.0149$  & 2012-10   & $184$ & $113$  & FTS \\
VII Zw 031  & \phn79.19333  & +79.67028 & $0.0540$  & 2010-06   & \phn$42$  & $137$  & WILMA \\
IRAS F05189-2524    & \phn80.25612  & \phantom{-}--25.36261 & $0.0435$  & 2011-12   & $297$ & $107$  & FTS \\
IRAS F05187-1017    & \phn80.27728  & \phantom{-}--10.24619 & $0.0283$  & 2011-12   & $138$ & \phn$88$   & FTS \\
IRAS F06076-2139    & \phn92.44088  & \phantom{-}--21.67325 & $0.0376$  & 2013-08   & $127$ & $107$  & WILMA \\
NGC 2341    & 107.30000 & +20.60278 & $0.0174$  & 2014-03   & \phn$61$ & \phn$89$   & FTS \\
NGC 2342    & 107.32525 & +20.63622 & $0.0174$  & 2012-10   & $326$ & $107$  & FTS \\
IRAS 07251-0248 & 111.90646 & \phn\phantom{-}--2.91503  & $0.0876$  & 2013-08   & $255$ & $113$  & WILMA \\
NGC 2623    & 129.60045 & +25.75461 & $0.0185$  & 2010-06   & \phn$42$  & $200$  & WILMA \\
IRAS 09111-1007W    & 138.40167 & \phantom{-}--10.32500 & $0.0564$  & 2013-08   & $318$ & $103$  & WILMA \\
IRAS 09111-1007E    & 138.41167 & \phantom{-}--10.32231 & $0.0550$  & 2011-09   & $117$ & \phn$98$   & WILMA \\
UGC 05101   & 143.96500 & +61.35328 & $0.0394$  & 2010-06   & $191$ & $116$  & WILMA \\
CGCG 011-076 & 170.30095 & \phn\phantom{-}--2.98396  & $0.0247$  & 2012-10   & $163$ & $123$  & FTS \\
IRAS F12224-0624    & 186.26621 & \phn\phantom{-}--6.68103  & $0.0257$  & 2012-10   & \phn$92$  & $135$  & FTS \\
CGCG 043-099 & 195.46167 & \phn+4.33333  & $0.0374$  & 2011-09   & $116$ & \phn$98$   & WILMA \\
ESO 507-G 070    & 195.71812 & \phantom{-}--23.92158 & $0.0215$  & 2011-12   & $180$ & $110$  & FTS \\
NGC 5104    & 200.34627 & \phn+0.34248  & $0.0186$  & 2012-10   & $153$ & $121$  & FTS \\
IC 4280 & 203.22212 & \phantom{-}--24.20720 & $0.0165$  & 2012-10   & $132$& $145$   & FTS \\
NGC 5257    & 204.97042 & \phn+0.84058  & $0.0227$  & 2014-03   & $132$    & $104$  & FTS \\
NGC 5258    & 204.98854 & \phn+0.82989  & $0.0226$  & 2011-12   & $180$ & \phn$88$   & FTS \\
UGC 08739   & 207.30800 & +35.25730 & $0.0172$  & 2012-10   & $102$ & $111$  & FTS \\
NGC 5331    & 208.06750 & \phn+2.10156  & $0.0330$  & 2013-08   & $286$ & \phn$98$   & WILMA \\
CGCG 247-020 & 214.93007 & +49.23666 & $0.0258$  & 2014-03   & $122$ & $153$  & FTS \\
IRAS F14348-1447    & 219.40975 & \phantom{-}--15.00633 & $0.0830$  & 2011-09   & $223$ & $139$  & WILMA \\
CGCG 049-057 & 228.30455 & \phn+7.22556  & $0.0127$  & 2012-10   & \phn$71$  & $116$  & FTS \\
NGC 5936    & 232.50360 & +12.98953 & $0.0134$  & 2012-10   & \phn$91$  & $123$  & FTS \\
ARP 220  & 233.73854 & +23.50323 & $0.0181$  & 2010-06   & \phn$53$  & $159$  & WILMA \\
IRAS F16164-0746    & 244.79913 & \phn\phantom{-}--7.90078  & $0.0229$  & 2011-12   & $201$ & \phn$94$   & FTS \\
CGCG 052-037 & 247.73545 & \phn+4.08292  & $0.0248$  & 2012-10   & $214$ & $113$  & FTS \\
IRAS F16399-0937    & 250.66754 & \phn\phantom{-}--9.72067  & $0.0270$  & 2011-12   & $191$ & \phn$91$   & FTS \\
NGC 6285    & 254.59998 & +58.95594 & $0.0186$  & 2011-12   & $127$ & \phn$82$   & FTS \\
NGC 6286    & 254.63146 & +58.93673 & $0.0185$  & 2011-12   & \phn$53$  & \phn$79$   & FTS \\
IRAS F17138-1017    & 259.14900 & \phantom{-}--10.34439 & $0.0175$  & 2011-12   & $149$ & $102$  & FTS \\
UGC 11041   & 268.71599 & +34.77625 & $0.0161$  & 2012-09   & \phn$41$  & $101$  & FTS \\
CGCG 141-034 & 269.23598 & +24.01704 & $0.0199$  & 2012-10   & \phn$61$  & $109$  & FTS \\
IRAS 18090+0130 & 272.91004 & \phn+1.52782  & $0.0293$  & 2013-08   & $286$ & \phn$94$   & WILMA \\
NGC 6701    & 280.80208 & +60.65312 & $0.0132$  & 2012-10   & $163$ & $165$  & FTS \\
NGC 6786    & 287.72500 & +73.40992 & $0.0250$  & 2011-09   & $265$ & \phn$95$   & WILMA \\
UGC 11415   & 287.76833 & +73.42556 & $0.0252$  & 2011-09   & $287$ & \phn$92$   & WILMA \\
ESO 593-IG 008   & 288.62950 & \phantom{-}--21.31897 & $0.0487$  & 2011-09   & \phn$53$  & $120$  & WILMA \\
NGC 6907    & 306.27750 & \phantom{-}--24.80893 & $0.0106$  & 2012-09   & $102$ & $171$  & FTS \\
IRAS 21101+5810 & 317.87667 & +58.38422 & $0.0398$  & 2013-08   & $308$ & \phn$89$   & WILMA \\
ESO 602-G 025    & 337.85621 & \phantom{-}--19.03454 & $0.0247$  & 2012-10   & $142$ & $125$  & FTS \\
UGC 12150   & 340.30077 & +34.24918 & $0.0216$  & 2012-10   & \phn$81$  & $100$  & FTS \\
IRAS F22491-1808    & 342.95567 & \phantom{-}--17.87357 & $0.0760$  & 2011-12   & $158$ & $111$  & FTS \\
CGCG 453-062 & 346.23565 & +19.55198 & $0.0248$  & 2012-10   & $122$ & \phn$96$   & FTS \\
NGC 7591    & 349.56777 & \phn+6.58579  & $0.0165$  & 2012-10   & $101$ & $102$  & FTS \\
IRAS F23365+3604    & 354.75542 & +36.35250 & $0.0645$  & 2011-09   & \phn$32$  & \phn$87$   & WILMA
\enddata
\label{table:obs}
\tablecomments{Col 1: Source name, Col 2: Right Ascension, Col 3: Declination, Col 4: redshift, Col 5: Year and month of observation, Col 6: Total on-source time, Col 7: System temperature, Col 8: Backend used for measurements (either the Fourier Transform Spectrometer \citep[FTS;][]{Klein2012} or the Wideband Line Multiple Autocorrelator (WILMA)).}
\end{deluxetable}

\begin{deluxetable}{lcccc}
\tablecaption{IRAM 30m Fluxes}
\tablehead{\colhead{Source} &
 \colhead{$\Sigma$ T$_{mb}dv$ (CCH)} &
 \colhead{$\Sigma$ T$_{mb}dv$ (\HCN{1}{0})} &
 \colhead{$\Sigma$ T$_{mb}dv$ (\HCO{1}{0})} &
 \colhead{$\Sigma$ T$_{mb}dv$ (\HNC{1}{0})} \\
    &
 \colhead{[K \kms]} &
 \colhead{[K \kms]} &
 \colhead{[K \kms]} &
 \colhead{[K \kms]}}
\startdata
NGC 0034    & $< 0.35$  & $0.71 \pm 0.12$   & $1.14 \pm 0.15$   & $< 0.44$ \\
MCG --02-01-052 & $< 0.33$  & $0.57 \pm 0.13$   & $< 0.27$  & $< 0.26$ \\
MCG --02-01-051 & $< 0.23$  & $0.17 \pm 0.05$   & $0.30 \pm 0.07$   & $< 0.16$ \\
IC 1623 & $1.40 \pm 0.14$   & $1.51 \pm 0.15$   & $4.00 \pm 0.15$   & $0.64 \pm 0.13$ \\
MCG --03-04-014 & $0.48 \pm 0.12$   & $0.94 \pm 0.12$   & $0.93 \pm 0.09$   & $0.38 \pm 0.12$ \\
IRAS 01364-1042 & $< 0.46$  & $< 0.71$  & $< 0.45$  & $< 0.54$ \\
IC 214  & $< 0.15$  & $0.36 \pm 0.10$   & $0.56 \pm 0.12$   & $< 0.14$ \\
NGC 0958    & $< 0.42$  & $< 0.71$  & $< 0.58$  & $< 0.69$ \\
ESO 550-I G025    & $< 0.35$  & $0.45 \pm 0.11$   & $0.36 \pm 0.11$   & $< 0.39$ \\
UGC 03094   & $0.63 \pm 0.12$   & $< 0.42$  & $0.74 \pm 0.14$   & $< 0.41$ \\
NGC 1797    & $< 0.37$  & $0.72 \pm 0.17$   & $0.55 \pm 0.12$   & $< 0.43$ \\
VII Zw 031  & $< 1.00$  & $< 0.98$  & $1.26 \pm 0.40$   & $< 0.96$ \\
IRAS F05189-2524    & $< 0.24$  & $0.73 \pm 0.13$   & $0.55 \pm 0.13$   & $0.39 \pm 0.11$ \\
IRAS F05187-1017    & $0.49 \pm 0.12$   & $0.75 \pm 0.11$   & $0.76 \pm 0.14$   & $< 0.18$ \\
IRAS F06076-2139    & $< 0.60$  & $1.05 \pm 0.20$   & $< 0.59$  & $< 0.50$ \\
NGC 2341    & $0.65 \pm 0.21$   & $< 0.35$  & $< 0.43$  & $< 0.60$ \\
NGC 2342    & $0.55 \pm 0.11$   & $0.68 \pm 0.11$   & $0.75 \pm 0.13$   & $< 0.22$ \\
IRAS 07251-0248 & $< 0.41$  & $0.35 \pm 0.11$   & $< 0.33$  & $< 0.40$ \\
NGC 2623    & $< 1.27$  & $2.62 \pm 0.48$   & $2.40 \pm 0.54$   & $< 1.23$ \\
IRAS 09111-1007W    & $< 0.36$  & $0.60 \pm 0.08$   & $0.83 \pm 0.16$   & $< 0.25$ \\
IRAS 09111-1007E    & $0.66 \pm 0.16$\tablenotemark{a}   & $< 0.36$  & $< 0.41$  & $< 0.35$ \\
UGC 05101   & $0.99 \pm 0.20$   & $2.17 \pm 0.20$   & $1.25 \pm 0.17$   & $0.92 \pm 0.20$ \\
CGCG 011-076 & $< 0.51$  & $1.09 \pm 0.15$   & $1.20 \pm 0.17$   & $< 0.35$ \\
IRAS F12224-0624    & $< 0.69$  & $< 0.55$  & $< 0.67$  & $< 0.66$ \\
CGCG 043-099 & $< 0.55$  & $< 0.54$  & $< 0.53$  & $< 0.53$ \\
ESO 507-G 070    & $< 0.48$  & $1.22 \pm 0.19$   & $1.87 \pm 0.21$   & $0.49 \pm 0.13$ \\
NGC 5104    & $< 0.67$  & $1.23 \pm 0.22$   & $0.74 \pm 0.20$   & $< 0.41$ \\
IC 4280 & $0.55 \pm 0.15$   & $0.95 \pm 0.21$   & $1.17 \pm 0.23$   & $0.57 \pm 0.18$ \\
NGC 5257    & $< 0.29$  & $0.39 \pm 0.10$   & $0.46 \pm 0.09$   & $0.37 \pm 0.11$ \\
NGC 5258    & $< 0.20$  & $0.43 \pm 0.10$   & $0.53 \pm 0.07$   & $0.30 \pm 0.08$ \\
UGC 08739   & $< 0.51$  & $1.30 \pm 0.17$   & $0.94 \pm 0.19$   & $0.88 \pm 0.19$ \\
NGC 5331    & $< 0.31$  & $0.67 \pm 0.10$   & $0.45 \pm 0.10$   & $< 0.43$ \\
CGCG 247-020 & $< 0.38$  & $0.96 \pm 0.15$   & $0.57 \pm 0.12$   & $0.35 \pm 0.09$ \\
IRAS F14348-1447    & $< 0.46$  & $< 0.52$  & $< 0.37$  & $< 0.51$ \\
CGCG 049-057 & $1.57 \pm 0.36$   & $3.21 \pm 0.32$   & $1.48 \pm 0.20$   & $1.70 \pm 0.28$ \\
NGC 5936    & $< 0.48$  & $1.21 \pm 0.16$   & $0.81 \pm 0.11$   & $< 0.46$ \\
ARP 220  & $4.37 \pm 0.62$   & $12.15 \pm 0.75$  & $6.02 \pm 0.74$   & $7.85 \pm 0.60$ \\
IRAS F16164-0746    & $0.65 \pm 0.12$   & $0.54 \pm 0.11$   & $0.93 \pm 0.12$   & $0.32 \pm 0.09$ \\
CGCG 052-037 & $0.57 \pm 0.12$   & $0.67 \pm 0.12$   & $0.74 \pm 0.10$   & $0.73 \pm 0.13$ \\
IRAS F16399-0937    & $0.66 \pm 0.14$   & $0.77 \pm 0.10$   & $0.68 \pm 0.10$   & $0.38 \pm 0.11$ \\
NGC 6285    & $0.38 \pm 0.10$   & $< 0.36$  & $< 0.25$  & $< 0.35$ \\
NGC 6286    & $0.93 \pm 0.20$   & $1.63 \pm 0.22$   & $1.67 \pm 0.21$   & $< 0.40$ \\
IRAS F17138-1017    & $1.01 \pm 0.17$   & $0.83 \pm 0.10$   & $1.03 \pm 0.12$   & $0.47 \pm 0.11$ \\
UGC 11041   & $< 0.50$  & $1.51 \pm 0.26$   & $1.49 \pm 0.20$   & $< 0.59$ \\
CGCG 141-034 & $< 0.51$  & $1.01 \pm 0.20$   & $< 0.49$  & $< 0.69$ \\
IRAS 18090+0130 & $< 0.30$  & $0.67 \pm 0.10$   & $0.78 \pm 0.11$   & $0.28 \pm 0.08$ \\
NGC 6701    & $< 0.48$  & $2.77 \pm 0.22$   & $2.26 \pm 0.22$   & $1.16 \pm 0.22$ \\
NGC 6786    & $0.30 \pm 0.10$   & $0.51 \pm 0.08$   & $0.56 \pm 0.11$   & $0.35 \pm 0.09$ \\
UGC 11415   & $< 0.36$  & $< 0.28$  & $0.54 \pm 0.12$   & $< 0.22$ \\
ESO 593-IG 008   & $< 0.77$  & $< 0.98$  & $< 1.22$  & $< 1.05$ \\
NGC 6907    & $< 0.83$  & $2.29 \pm 0.31$   & $1.27 \pm 0.27$   & $< 0.65$ \\
IRAS 21101+5810 & $0.47 \pm 0.11$   & $0.53 \pm 0.09$   & $0.33 \pm 0.08$   & $< 0.22$ \\
ESO 602-G 025    & $< 0.57$  & $0.84 \pm 0.21$   & $0.75 \pm 0.12$   & $0.74 \pm 0.20$ \\
UGC 12150   & $0.69 \pm 0.15$   & $1.36 \pm 0.17$   & $1.59 \pm 0.14$   & $0.51 \pm 0.14$ \\
IRAS F22491-1808    & $< 0.83$  & $< 0.82$  & $< 0.73$  & $< 0.71$ \\
CGCG 453-062 & $0.74 \pm 0.14$   & $0.97 \pm 0.16$   & $0.48 \pm 0.16$   & $< 0.42$ \\
NGC 7591    & $1.37 \pm 0.20$   & $1.35 \pm 0.18$   & $0.69 \pm 0.13$   & $0.68 \pm 0.20$ \\
IRAS F23365+3604    & $< 0.58$  & $0.81 \pm 0.27$   & $1.28 \pm 0.27$   & $< 0.88$
\enddata
\label{table:fluxes}
\tablecomments{Col 1: Source name, Col 2: \CCHn{1}{0} flux with $1\sigma$ errors, Col 3: \HCNj{1}{0} flux with $1\sigma$ errors, Col 4: \HCOj{1}{0} flux with $1\sigma$ errors, Col 5: \HNCj{1}{0} flux with $1\sigma$ errors.
The quoted upper limits are $3\sigma$.}
\tablenotetext{a}{Though formally a detection ($>3\sigma$), this CCH flux appears to be spurious as the signal at the location of CCH is significantly broader than would be expected.}
\end{deluxetable}

\begin{deluxetable}{lcc}
\tablecaption{Spearman Rank Coefficients}
\tablehead{\colhead{Quantities} &
 \colhead{$\rho$} &
 \colhead{$z$-score}}
\startdata
\Lprime{HCN}{1}{0}/\Lprime{HCO$^+$}{1}{0} vs $\log_{10}$(\LIR/\Lsun) (Figure~\ref{fig:LIR}) & \phn$0.021 \pm 0.178$ & \phn\phantom{-}$0.057 \pm 0.494 $ \\
\Lprime{HCN}{1}{0}/\Lprime{HCO$^+$}{1}{0} vs $6.2~\mu m$ PAH EQW (Figure~\ref{fig:AGNtracer},~Left) & $-0.512 \pm 0.127$ & $-1.532\pm0.464$\\
\Lprime{HCN}{1}{0}/\Lprime{HCO$^+$}{1}{0} vs $\log_{10}$(\cii/\LFIR) (Figure~\ref{fig:deficit}) & $-0.401\pm0.142$ & $-1.165 \pm 0.458$ 
\enddata
\label{table:spearman}
\tablecomments{Spearman rank coefficients for several relationships discussed in the text along with $z$-scores.
Quantities and their 68\% confidence intervals were computed according to the Monte Carlo perturbation plus bootstrapping method (1e5 iterations) discussed by \citet{Curran2014} and using the associated code \citep{Curran2015}.
Upper and lower limits for \Lprime{HCN}{1}{0}/\Lprime{HCO$^+$}{1}{0} were not included when computing these coefficients.}
\end{deluxetable}

\begin{figure*}
\includegraphics[width=\textwidth]{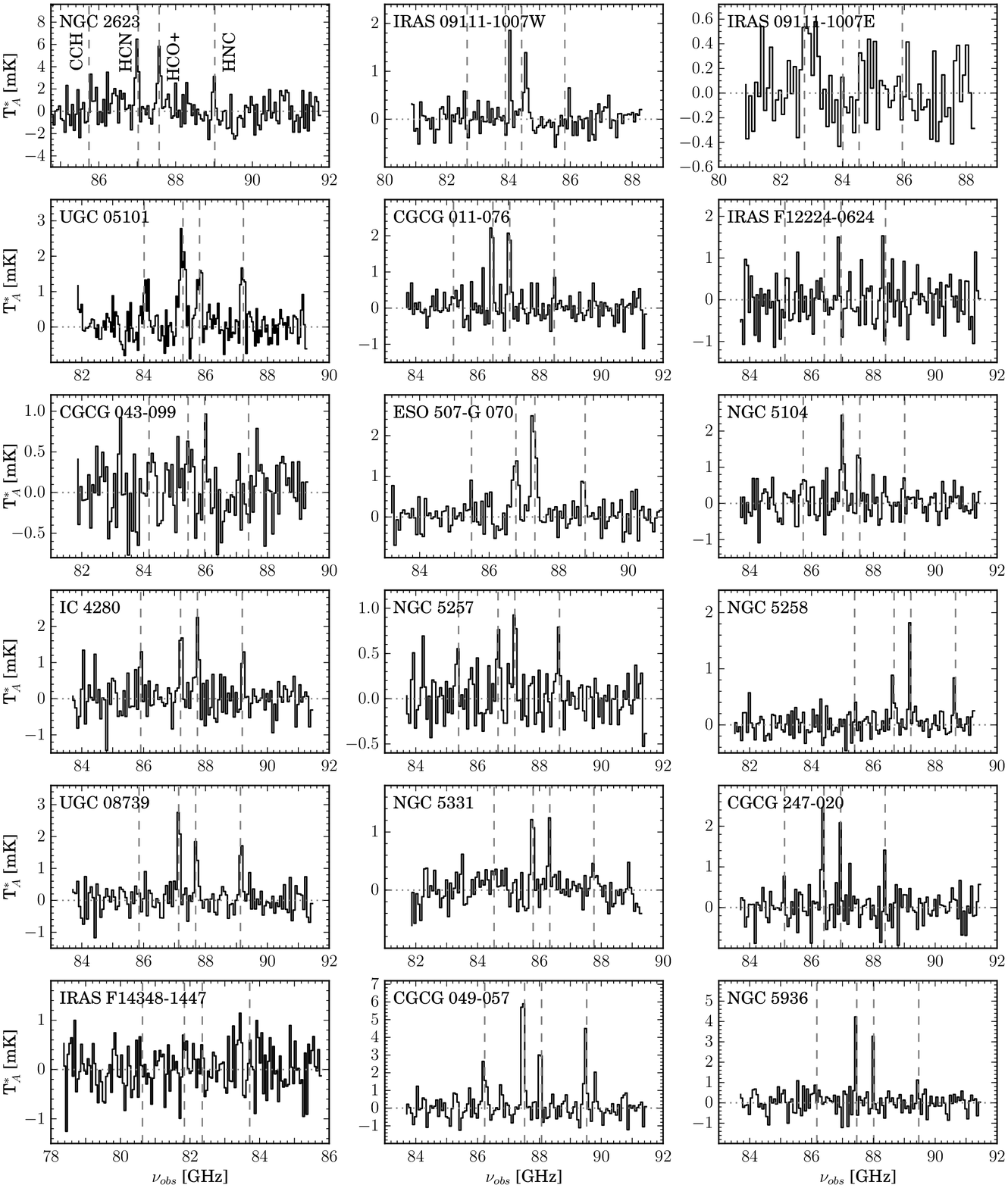}
\caption{Figure~\ref{fig:spectra} Continued}
\end{figure*}

\begin{figure*}
\includegraphics[width=\textwidth]{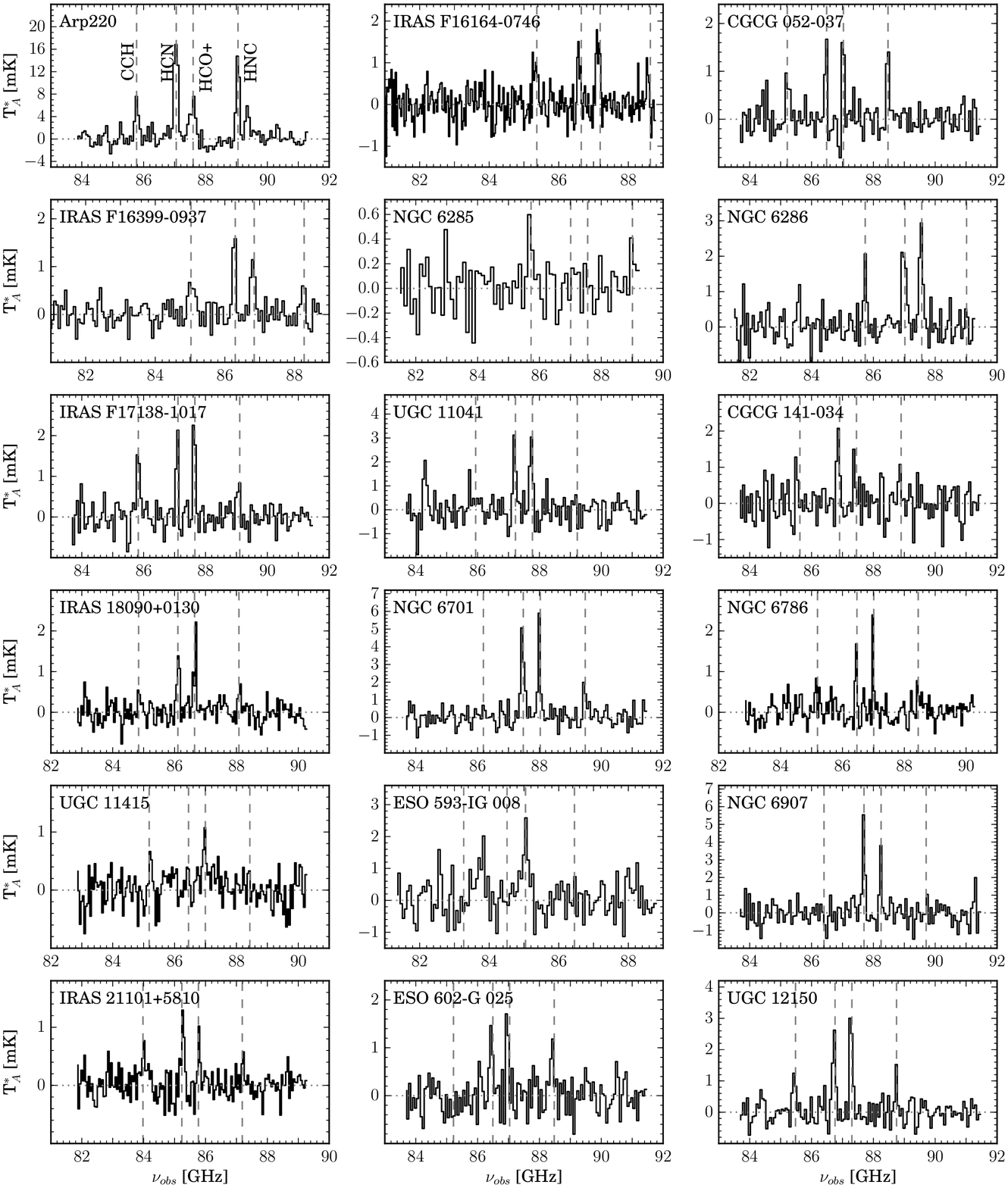}
\caption{Figure~\ref{fig:spectra} Continued}
\end{figure*}

\begin{figure*}
\includegraphics[width=\textwidth]{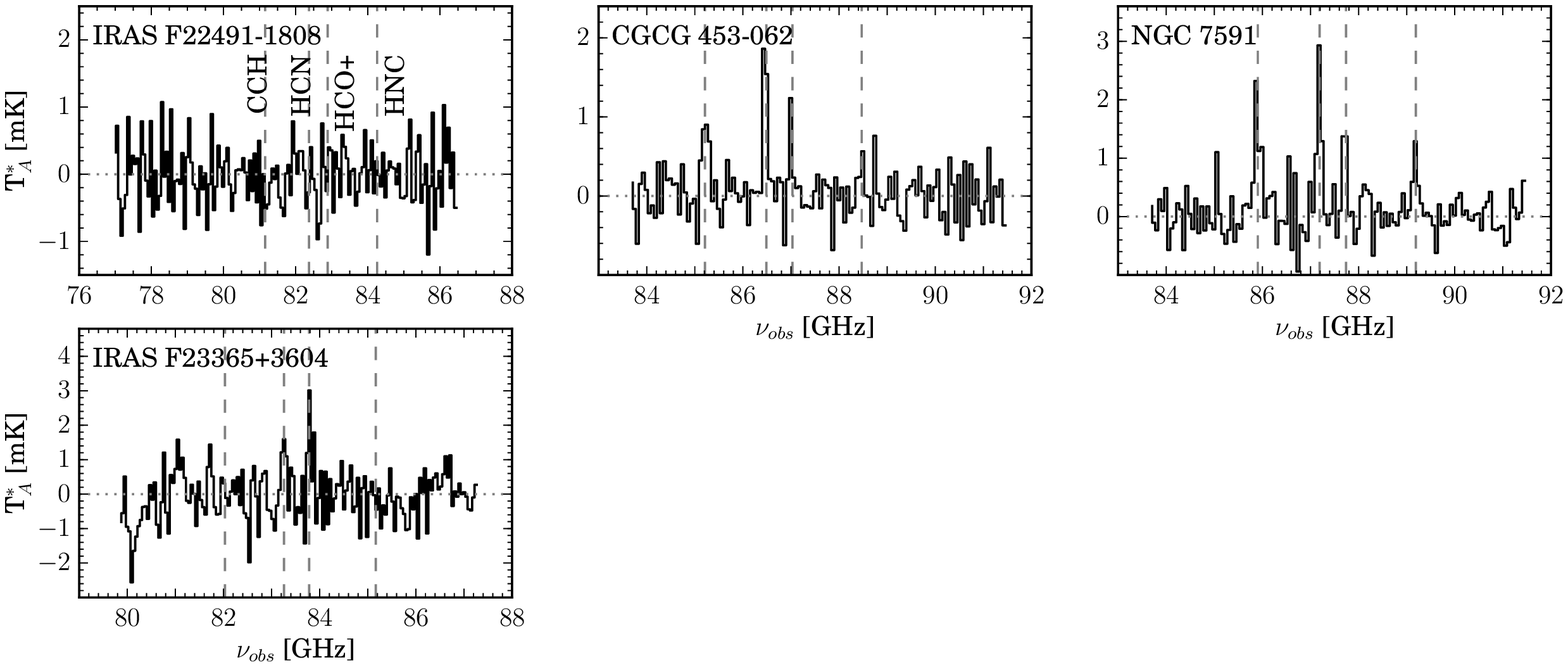}
\caption{Figure~\ref{fig:spectra} Continued}
\end{figure*}


\begin{thebibliography}{59}
\expandafter\ifx\csname natexlab\endcsname\relax\def\natexlab#1{#1}\fi

\bibitem[{{Aalto} {et~al.}(1995){Aalto}, {Booth}, {Black}, \&
  {Johansson}}]{Aalto1995}
{Aalto}, S., {Booth}, R.~S., {Black}, J.~H., \& {Johansson}, L.~E.~B. 1995,
  \aap, 300, 369

\bibitem[{{Aalto} {et~al.}(1994){Aalto}, {Booth}, {Black}, {Koribalski}, \&
  {Wielebinski}}]{Aalto1994}
{Aalto}, S., {Booth}, R.~S., {Black}, J.~H., {Koribalski}, B., \&
  {Wielebinski}, R. 1994, \aap, 286, 365

\bibitem[{{Aalto} {et~al.}(2007){Aalto}, {Spaans}, {Wiedner}, \&
  {H{\"u}ttemeister}}]{Aalto2007}
{Aalto}, S., {Spaans}, M., {Wiedner}, M.~C., \& {H{\"u}ttemeister}, S. 2007,
  \aap, 464, 193

\bibitem[{{Aalto} {et~al.}(2015){Aalto}, {Costagliola}, {Gonzalez-Alfonso},
  {Muller}, {Sakamoto}, {Fuller}, {Garcia-Burillo}, {van der Werf}, {Neri},
  {Spaans}, {Combes}, {Viti}, {Muehle}, {Armus}, {Evans}, {Sturm},
  {Cernicharo}, {Henkel}, \& {Greve}}]{Aalto2015a}
{Aalto}, S., {et~al.} 2015, ArXiv e-prints, astro-ph:1504.06824

\bibitem[{{Armus} {et~al.}(2004){Armus}, {Charmandaris}, {Spoon}, {Houck},
  {Soifer}, {Brandl}, {Appleton}, {Teplitz}, {Higdon}, {Weedman}, {Devost},
  {Morris}, {Uchida}, {van Cleve}, {Barry}, {Sloan}, {Grillmair}, {Burgdorf},
  {Fajardo-Acosta}, {Ingalls}, {Higdon}, {Hao}, {Bernard-Salas}, {Herter},
  {Troeltzsch}, {Unruh}, \& {Winghart}}]{Armus2004}
{Armus}, L., {et~al.} 2004, \apjs, 154, 178

\bibitem[{{Armus} {et~al.}(2007){Armus}, {Charmandaris}, {Bernard-Salas},
  {Spoon}, {Marshall}, {Higdon}, {Desai}, {Teplitz}, {Hao}, {Devost}, {Brandl},
  {Wu}, {Sloan}, {Soifer}, {Houck}, \& {Herter}}]{Armus2007}
---. 2007, \apj, 656, 148

\bibitem[{Armus {et~al.}(2009)Armus, Mazzarella, Evans, Surace, Sanders,
  Iwasawa, Frayer, Howell, Chan, Petric, Vavilkin, Kim, Haan, Inami, Murphy,
  Appleton, Barnes, Bothun, Bridge, Charmandaris, Jensen, Kewley, Lord, Madore,
  Marshall, Melbourne, Rich, Satyapal, Schulz, Spoon, Sturm, U, Veilleux, \&
  Xu}]{Armus2009}
Armus, L., {et~al.} 2009, \pasp, 121, 559

\bibitem[{{Barcos-Mu{\~n}oz} {et~al.}(2015){Barcos-Mu{\~n}oz}, {Leroy},
  {Evans}, {Privon}, {Armus}, {Condon}, {Mazzarella}, {Meier}, {Momjian},
  {Murphy}, {Ott}, {Reichardt}, {Sakamoto}, {Sanders}, {Schinnerer},
  {Stierwalt}, {Surace}, {Thompson}, \& {Walter}}]{Barcos-Munoz2015}
{Barcos-Mu{\~n}oz}, L., {et~al.} 2015, \apj, 799, 10

\bibitem[{Bigiel {et~al.}(2008)Bigiel, Leroy, Walter, Brinks, de~Blok, Madore,
  \& Thornley}]{Bigiel2008}
Bigiel, F., Leroy, A., Walter, F., Brinks, E., de~Blok, W. J.~G., Madore, B.,
  \& Thornley, M.~D. 2008, \aj, 136, 2846

\bibitem[{{Brandl} {et~al.}(2006){Brandl}, {Bernard-Salas}, {Spoon}, {Devost},
  {Sloan}, {Guilles}, {Wu}, {Houck}, {Weedman}, {Armus}, {Appleton}, {Soifer},
  {Charmandaris}, {Hao}, {Higdon}, {Marshall}, \& {Herter}}]{Brandl2006}
{Brandl}, B.~R., {et~al.} 2006, \apj, 653, 1129

\bibitem[{{Carter} {et~al.}(2012){Carter}, {Lazareff}, {Maier}, {Chenu},
  {Fontana}, {Bortolotti}, {Boucher}, {Navarrini}, {Blanchet}, {Greve}, {John},
  {Kramer}, {Morel}, {Navarro}, {Pe{\~n}alver}, {Schuster}, \&
  {Thum}}]{Carter2012}
{Carter}, M., {et~al.} 2012, \aap, 538, A89

\bibitem[{Costagliola {et~al.}(2011)Costagliola, Aalto, Rodriguez, Muller,
  Spoon, Mart\'{\i}n, Per\'{e}z-Torres, Alberdi, Lindberg, Batejat, J\"{u}tte,
  {van Der Werf}, \& Lahuis}]{Costagliola2011}
Costagliola, F., {et~al.} 2011, \aap, 528, A30

\bibitem[{{Curran}(2014)}]{Curran2014}
{Curran}, P.~A. 2014, ArXiv e-prints, astro-ph:1411.3816

\bibitem[{{Curran}(2015)}]{Curran2015}
---. 2015

\bibitem[{Davies {et~al.}(2012)Davies, Mark, \& Sternberg}]{Davies2012}
Davies, R.~I., Mark, D., \& Sternberg, A. 2012, \aap, 537, A133

\bibitem[{{Diaz-Santos} {et~al.}(2013){Diaz-Santos}, {Armus}, {Charmandaris},
  {Stierwalt}, {Murphy}, {Haan}, {Inami}, {Malhotra}, {Meijerink}, {Stacey},
  {Petric}, {Evans}, {Veilleux}, {van der Werf}, {Lord}, {Lu}, {Howell},
  {Appleton}, {Mazzarella}, {Surace}, {Xu}, {Schulz}, {Sanders}, {Bridge},
  {Chan}, {Frayer}, {Iwasawa}, {Melbourne}, \& {Sturm}}]{Diaz-Santos2013}
{Diaz-Santos}, T., {et~al.} 2013, \apj, 774, 68

\bibitem[{{Diaz-Santos} {et~al.}(2014){Diaz-Santos}, {Armus}, {Charmandaris},
  {Stacey}, {Murphy}, {Haan}, {Stierwalt}, {Malhotra}, {Appleton}, {Inami},
  {Magdis}, {Elbaz}, {Evans}, {Mazzarella}, {Surace}, {van der Werf}, {Xu},
  {Lu}, {Meijerink}, {Howell}, {Petric}, {Veilleux}, \&
  {Sanders}}]{Diaz-Santos2014}
---. 2014, \apj, 788, L17

\bibitem[{{Evans} {et~al.}(2006){Evans}, {Solomon}, {Tacconi}, {Vavilkin}, \&
  {Downes}}]{Evans2006}
{Evans}, A.~S., {Solomon}, P.~M., {Tacconi}, L.~J., {Vavilkin}, T., \&
  {Downes}, D. 2006, \aj, 132, 2398

\bibitem[{{Foreman-Mackey} {et~al.}(2013){Foreman-Mackey}, {Hogg}, {Lang}, \&
  {Goodman}}]{Foreman-Mackey2013}
{Foreman-Mackey}, D., {Hogg}, D.~W., {Lang}, D., \& {Goodman}, J. 2013, \pasp,
  125, 306

\bibitem[{Gao \& Solomon(2004{\natexlab{a}})}]{Gao2004}
Gao, Y., \& Solomon, P.~M. 2004{\natexlab{a}}, \apjs, 152, 63

\bibitem[{Gao \& Solomon(2004{\natexlab{b}})}]{Gao2004a}
---. 2004{\natexlab{b}}, \apj, 606, 271

\bibitem[{Garc\'{\i}a-Burillo {et~al.}(2012)Garc\'{\i}a-Burillo, Usero,
  Alonso-Herrero, Graci\'{a}-Carpio, Pereira-Santaella, Colina, Planesas, \&
  Arribas}]{GarciaBurillo2012}
Garc\'{\i}a-Burillo, S., Usero, A., Alonso-Herrero, A., Graci\'{a}-Carpio, J.,
  Pereira-Santaella, M., Colina, L., Planesas, P., \& Arribas, S. 2012, \aap,
  539, A8

\bibitem[{{Genzel} {et~al.}(1998){Genzel}, {Lutz}, {Sturm}, {Egami}, {Kunze},
  {Moorwood}, {Rigopoulou}, {Spoon}, {Sternberg}, {Tacconi-Garman}, {Tacconi},
  \& {Thatte}}]{Genzel1998}
{Genzel}, R., {et~al.} 1998, \apj, 498, 579

\bibitem[{{Ginsburg} \& {Mirocha}(2011)}]{Ginsburg2011c}
{Ginsburg}, A., \& {Mirocha}, J. 2011, Astrophysics Source Code Library, 9001

\bibitem[{Graci\'{a}-Carpio {et~al.}(2006)Graci\'{a}-Carpio,
  Garc\'{\i}a-Burillo, Planesas, \& Colina}]{GraciaCarpio2006}
Graci\'{a}-Carpio, J., Garc\'{\i}a-Burillo, S., Planesas, P., \& Colina, L.
  2006, \apj, 640, L135

\bibitem[{Haan {et~al.}(2011)Haan, Surace, Armus, Evans, Howell, Mazzarella,
  Kim, Vavilkin, Inami, Sanders, Petric, Bridge, Melbourne, Charmandaris,
  D\'{\i}az-Santos, Murphy, U, Stierwalt, \& Marshall}]{Haan2011}
Haan, S., {et~al.} 2011, \aj, 141, 100

\bibitem[{{Harrison} {et~al.}(2013){Harrison}, {Craig}, {Christensen},
  {Hailey}, {Zhang}, {Boggs}, {Stern}, {Cook}, {Forster}, {Giommi},
  {Grefenstette}, {Kim}, {Kitaguchi}, {Koglin}, {Madsen}, {Mao}, {Miyasaka},
  {Mori}, {Perri}, {Pivovaroff}, {Puccetti}, {Rana}, {Westergaard}, {Willis},
  {Zoglauer}, {An}, {Bachetti}, {Barri{\`e}re}, {Bellm}, {Bhalerao},
  {Brejnholt}, {Fuerst}, {Liebe}, {Markwardt}, {Nynka}, {Vogel}, {Walton},
  {Wik}, {Alexander}, {Cominsky}, {Hornschemeier}, {Hornstrup}, {Kaspi},
  {Madejski}, {Matt}, {Molendi}, {Smith}, {Tomsick}, {Ajello}, {Ballantyne},
  {Balokovi{\'c}}, {Barret}, {Bauer}, {Blandford}, {Brandt}, {Brenneman},
  {Chiang}, {Chakrabarty}, {Chenevez}, {Comastri}, {Dufour}, {Elvis}, {Fabian},
  {Farrah}, {Fryer}, {Gotthelf}, {Grindlay}, {Helfand}, {Krivonos}, {Meier},
  {Miller}, {Natalucci}, {Ogle}, {Ofek}, {Ptak}, {Reynolds}, {Rigby},
  {Tagliaferri}, {Thorsett}, {Treister}, \& {Urry}}]{Harrison2013}
{Harrison}, F.~A., {et~al.} 2013, \apj, 770, 103

\bibitem[{{Hinshaw} {et~al.}(2009){Hinshaw}, {Weiland}, {Hill}, {Odegard},
  {Larson}, {Bennett}, {Dunkley}, {Gold}, {Greason}, {Jarosik}, {Komatsu},
  {Nolta}, {Page}, {Spergel}, {Wollack}, {Halpern}, {Kogut}, {Limon}, {Meyer},
  {Tucker}, \& {Wright}}]{Hinshaw2009}
{Hinshaw}, G., {et~al.} 2009, \apjs, 180, 225

\bibitem[{Hogg {et~al.}(2010)Hogg, Bovy, \& Lang}]{Hogg2010}
Hogg, D.~W., Bovy, J., \& Lang, D. 2010, ArXiv e-prints, astro-ph:1008.4686

\bibitem[{{Houck} {et~al.}(2004){Houck}, {Roellig}, {van Cleve}, {Forrest},
  {Herter}, {Lawrence}, {Matthews}, {Reitsema}, {Soifer}, {Watson}, {Weedman},
  {Huisjen}, {Troeltzsch}, {Barry}, {Bernard-Salas}, {Blacken}, {Brandl},
  {Charmandaris}, {Devost}, {Gull}, {Hall}, {Henderson}, {Higdon}, {Pirger},
  {Schoenwald}, {Sloan}, {Uchida}, {Appleton}, {Armus}, {Burgdorf},
  {Fajardo-Acosta}, {Grillmair}, {Ingalls}, {Morris}, \& {Teplitz}}]{Houck2004}
{Houck}, J.~R., {et~al.} 2004, \apjs, 154, 18

\bibitem[{Howell {et~al.}(2010)Howell, Armus, Mazzarella, Evans, Surace,
  Sanders, Petric, Appleton, Bothun, Bridge, Chan, Charmandaris, Frayer, Haan,
  Inami, Kim, Lord, Madore, Melbourne, Schulz, U, Vavilkin, Veilleux, \&
  Xu}]{Howell2010}
Howell, J.~H., {et~al.} 2010, \apj, 715, 572

\bibitem[{{Huettemeister} {et~al.}(1995){Huettemeister}, {Henkel},
  {Mauersberger}, {Brouillet}, {Wiklind}, \& {Millar}}]{Huettemeister1995}
{Huettemeister}, S., {Henkel}, C., {Mauersberger}, R., {Brouillet}, N.,
  {Wiklind}, T., \& {Millar}, T.~J. 1995, \aap, 295, 571

\bibitem[{{Imanishi} \& {Nakanishi}(2006)}]{Imanishi2006a}
{Imanishi}, M., \& {Nakanishi}, K. 2006, \pasj, 58, 813

\bibitem[{{Imanishi} \& {Nakanishi}(2013)}]{Imanishi2013}
---. 2013, \aj, 146, 91

\bibitem[{{Imanishi} \& {Nakanishi}(2014)}]{Imanishi2014a}
---. 2014, \aj, 148, 9

\bibitem[{{Imanishi} {et~al.}(2006){Imanishi}, {Nakanishi}, \&
  {Kohno}}]{Imanishi2006}
{Imanishi}, M., {Nakanishi}, K., \& {Kohno}, K. 2006, \aj, 131, 2888

\bibitem[{{Imanishi} {et~al.}(2007){Imanishi}, {Nakanishi}, {Tamura}, {Oi}, \&
  {Kohno}}]{Imanishi2007}
{Imanishi}, M., {Nakanishi}, K., {Tamura}, Y., {Oi}, N., \& {Kohno}, K. 2007,
  \aj, 134, 2366

\bibitem[{{Imanishi} {et~al.}(2009){Imanishi}, {Nakanishi}, {Tamura}, \&
  {Peng}}]{Imanishi2009}
{Imanishi}, M., {Nakanishi}, K., {Tamura}, Y., \& {Peng}, C.-H. 2009, \aj, 137,
  3581

\bibitem[{Inami {et~al.}(2013)Inami, Armus, Charmandaris, Groves, Kewley,
  Petric, Stierwalt, Díaz-Santos, Surace, Rich, Haan, Howell, Evans,
  Mazzarella, Marshall, Appleton, Lord, Spoon, Frayer, Matsuhara, \&
  Veilleux}]{Inami2013}
Inami, H., {et~al.} 2013, \apj, 777, 156

\bibitem[{Iwasawa {et~al.}(2011)Iwasawa, Mazzarella, Surace, Sanders, Armus,
  Evans, Howell, Komossa, Petric, Teng, U, \& Veilleux}]{Iwasawa2011}
Iwasawa, K., {et~al.} 2011, \aap, 528, A137

\bibitem[{{Juneau} {et~al.}(2009){Juneau}, {Narayanan}, {Moustakas}, {Shirley},
  {Bussmann}, {Kennicutt}, \& {Vanden Bout}}]{Juneau2009}
{Juneau}, S., {Narayanan}, D.~T., {Moustakas}, J., {Shirley}, Y.~L.,
  {Bussmann}, R.~S., {Kennicutt}, Jr., R.~C., \& {Vanden Bout}, P.~A. 2009,
  \apj, 707, 1217

\bibitem[{{Kim} {et~al.}(2013){Kim}, {Evans}, {Vavilkin}, {Armus},
  {Mazzarella}, {Sheth}, {Surace}, {Haan}, {Howell}, {D{\'{\i}}az-Santos},
  {Petric}, {Iwasawa}, {Privon}, \& {Sanders}}]{Kim2013}
{Kim}, D.-C., {et~al.} 2013, \apj, 768, 102

\bibitem[{{Klein} {et~al.}(2012){Klein}, {Hochg{\"u}rtel}, {Kr{\"a}mer},
  {Bell}, {Meyer}, \& {G{\"u}sten}}]{Klein2012}
{Klein}, B., {Hochg{\"u}rtel}, S., {Kr{\"a}mer}, I., {Bell}, A., {Meyer}, K.,
  \& {G{\"u}sten}, R. 2012, \aap, 542, L3

\bibitem[{{Kohno} {et~al.}(2003){Kohno}, {Ishizuki}, {Matsushita},
  {Vila-Vilar{\'o}}, \& {Kawabe}}]{Kohno2003}
{Kohno}, K., {Ishizuki}, S., {Matsushita}, S., {Vila-Vilar{\'o}}, B., \&
  {Kawabe}, R. 2003, \pasj, 55, L1

\bibitem[{Krips {et~al.}(2008)Krips, Neri, Garc\'{\i}a‚ÄêBurillo,
  Mart\'{\i}n, Combes, Graci\'{a}‚ÄêCarpio, \& Eckart}]{Krips2008}
Krips, M., Neri, R., Garc\'{\i}a‚ÄêBurillo, S., Mart\'{\i}n, S., Combes,
  F., Graci\'{a}‚ÄêCarpio, J., \& Eckart, A. 2008, \apj, 677, 262

\bibitem[{{Lahuis} {et~al.}(2007){Lahuis}, {Spoon}, {Tielens}, {Doty}, {Armus},
  {Charmandaris}, {Houck}, {St{\"a}uber}, \& {van Dishoeck}}]{Lahuis2007}
{Lahuis}, F., {et~al.} 2007, \apj, 659, 296

\bibitem[{{Lepp} \& {Dalgarno}(1996)}]{Lepp1996}
{Lepp}, S., \& {Dalgarno}, A. 1996, \aap, 306, L21

\bibitem[{Leroy {et~al.}(2011)Leroy, Evans, Momjian, Murphy, Ott, Armus,
  Condon, Haan, Mazzarella, Meier, Privon, Schinnerer, Surace, \&
  Walter}]{Leroy2011}
Leroy, A.~K., {et~al.} 2011, \apj, 739, L25

\bibitem[{{Leroy} {et~al.}(2012){Leroy}, {Bigiel}, {de Blok}, {Boissier},
  {Bolatto}, {Brinks}, {Madore}, {Munoz-Mateos}, {Murphy}, {Sandstrom},
  {Schruba}, \& {Walter}}]{Leroy2012}
{Leroy}, A.~K., {et~al.} 2012, \aj, 144, 3

\bibitem[{{Lu} {et~al.}(2014){Lu}, {Zhao}, {Xu}, {Gao}, {Armus}, {Mazzarella},
  {Isaak}, {Petric}, {Charmandaris}, {D{\'{\i}}az-Santos}, {Evans}, {Howell},
  {Appleton}, {Inami}, {Iwasawa}, {Leech}, {Lord}, {Sanders}, {Schulz},
  {Surace}, \& {van der Werf}}]{Lu2014}
{Lu}, N., {et~al.} 2014, \apjl, 787, L23

\bibitem[{{Mart{\'{\i}}n} {et~al.}(2014){Mart{\'{\i}}n}, {Verdes-Montenegro},
  {Aladro}, {Espada}, {Argudo-Fern{\'a}ndez}, {Kramer}, \&
  {Scott}}]{Martin2014}
{Mart{\'{\i}}n}, S., {Verdes-Montenegro}, L., {Aladro}, R., {Espada}, D.,
  {Argudo-Fern{\'a}ndez}, M., {Kramer}, C., \& {Scott}, T.~C. 2014, \aap, 563,
  L6

\bibitem[{{Meijerink} {et~al.}(2007){Meijerink}, {Spaans}, \&
  {Israel}}]{Meijerink2007}
{Meijerink}, R., {Spaans}, M., \& {Israel}, F.~P. 2007, \aap, 461, 793

\bibitem[{{Mould} {et~al.}(2000){Mould}, {Huchra}, {Freedman}, {Kennicutt},
  {Ferrarese}, {Ford}, {Gibson}, {Graham}, {Hughes}, {Illingworth}, {Kelson},
  {Macri}, {Madore}, {Sakai}, {Sebo}, {Silbermann}, \& {Stetson}}]{Mould2000}
{Mould}, J.~R., {et~al.} 2000, \apj, 529, 786

\bibitem[{Petric {et~al.}(2011)Petric, Armus, Howell, Chan, Mazzarella, Evans,
  Surace, Sanders, Appleton, Charmandaris, D\'{\i}az-Santos, Frayer, Haan,
  Inami, Iwasawa, Kim, Madore, Marshall, Spoon, Stierwalt, Sturm, U, Vavilkin,
  Veilleux, Santos, \& Lord}]{Petric2011}
Petric, A.~O., {et~al.} 2011, \apj, 730, 28

\bibitem[{Sakamoto {et~al.}(2010)Sakamoto, Aalto, Evans, Wiedner, \&
  Wilner}]{Sakamoto2010}
Sakamoto, K., Aalto, S., Evans, A.~S., Wiedner, M.~C., \& Wilner, D.~J. 2010,
  \apj, 725, 8

\bibitem[{Sanders {et~al.}(2003)Sanders, Mazzarella, Kim, Surace, \&
  Soifer}]{Sanders2003}
Sanders, D.~B., Mazzarella, J.~M., Kim, D.-C., Surace, J.~A., \& Soifer, B.~T.
  2003, \aj, 126, 1607

\bibitem[{Solomon {et~al.}(1992)Solomon, Radford, \& Downes}]{Solomon1992}
Solomon, P.~M., Radford, S. J.~E., \& Downes, D. 1992, \nat, 356, 318

\bibitem[{Stierwalt {et~al.}(2013)Stierwalt, Armus, Surace, Inami, Petric,
  D\'{\i}az-Santos, Haan, Charmandaris, Howell, Kim, Marshall, Mazzarella,
  Spoon, Veilleux, Evans, Sanders, Appleton, Bothun, Bridge, Chan, Frayer,
  Iwasawa, Kewley, Lord, Madore, Melbourne, Murphy, Rich, Schulz, Sturm, U,
  Vavilkin, \& Xu}]{Stierwalt2013}
Stierwalt, S., {et~al.} 2013, \apj, 206, 1

\bibitem[{{Teng} {et~al.}(2014){Teng}, {Brandt}, {Harrison}, {Luo},
  {Alexander}, {Bauer}, {Boggs}, {Christensen}, {Comastri}, {Craig}, {Fabian},
  {Farrah}, {Fiore}, {Gandhi}, {Grefenstette}, {Hailey}, {Hickox}, {Madsen},
  {Ptak}, {Rigby}, {Risaliti}, {Saez}, {Stern}, {Veilleux}, {Walton}, {Wik}, \&
  {Zhang}}]{Teng2014}
{Teng}, S.~H., {et~al.} 2014, \apj, 785, 19

\bibitem[{{The Astropy Collaboration} {et~al.}(2013){The Astropy
  Collaboration}, {Robitaille}, {Tollerud}, {Greenfield}, {Droettboom}, {Bray},
  {Aldcroft}, {Davis}, {Ginsburg}, {Price-Whelan}, {Kerzendorf}, {Conley},
  {Crighton}, {Barbary}, {Muna}, {Ferguson}, {Grollier}, {Parikh}, {Nair},
  {G{\"u}nther}, {Deil}, {Woillez}, {Conseil}, {Kramer}, {Turner}, {Singer},
  {Fox}, {Weaver}, {Zabalza}, {Edwards}, {Azalee Bostroem}, {Burke}, {Casey},
  {Crawford}, {Dencheva}, {Ely}, {Jenness}, {Labrie}, {Lian Lim},
  {Pierfederici}, {Pontzen}, {Ptak}, {Refsdal}, {Servillat}, \&
  {Streicher}}]{Astropy2013}
{The Astropy Collaboration} {et~al.} 2013, \aap, 558, A33

\end{thebibliography}
\end{document}